\documentstyle[12pt,epsfig]{article}
\makeatletter
\def\slash#1{{\mathpalette\c@ncel{#1}}} 
\makeatother

\newcommand\beq{\begin{eqnarray}}
\newcommand\eeq{\end{eqnarray}}

\newcommand\la{\langle}
\newcommand\ra{\rangle}

\newcommand{\sslash}[1]{\not{\!#1}}
\newcommand{\lslash}[1]{\not{\!\!#1}}

\def\ellslash{\rlap/{\mkern-1mu \ell}}

\def\pslash{\rlap/{\mkern-1mu p}}

\def\wslash{\rlap/{\mkern-1mu w}}
\def\ellslash{\rlap/{\mkern-1mu \ell}}

\def\nslash{\slash{\mkern-1mu n}}

\def\xhat{\widehat{x}}
\def\zhat{\widehat{z}}

\def\pvec{\vec{p}}
\def\qvec{\vec{q}}
\def\kvec{\vec{k}}

\def\Svec{\vec{S}}

\def\GFt{\widetilde{G}_F}
\def\GDt{\widetilde{G}_D}
\def\gt{\widetilde{g}}

\begin{document}
\begin{flushright}
\end{flushright}
\vspace*{15mm}
\begin{center}
{\Large \bf Single Transverse Spin Asymmetry \\[4mm]
for Large-$p_T$ Pion Production in \\[6mm] Semi-Inclusive Deep Inelastic Scattering}
\vspace{1.5cm}\\
 {\sc Hisato Eguchi$^1$, Yuji Koike$^1$, Kazuhiro Tanaka$^2$}
\\[0.4cm]
\vspace*{0.1cm}{\it $^1$ Department of Physics, Niigata University,
Ikarashi, Niigata 950-2181, Japan}\\
\vspace*{0.1cm}{\it $^2$ Department of Physics, 
Juntendo University, Inba-gun, Chiba 270-1695, Japan}
\\[3cm]

{\large \bf Abstract} \end{center}
We study the single spin asymmetry (SSA) for the
pion production with large transverse momentum $p_T$ in 
semi-inclusive deep inelastic scattering $ep^\uparrow\to e\pi X$.  
We derive the twist-3 cross section formula for SSA, focussing on the
soft-gluon-pole contributions associated 
with the twist-3 distribution for the nucleon and with the
twist-3 fragmentation function for the pion.
We present a simple estimate of the asymmetries due to 
each twist-3 effect from nucleon and pion, respectively,
by fixing the overall
strength of the relevant nonperturbative quantities by the data on the SSA $A_N$ 
in $p^\uparrow p\to\pi X$ collision.

\noindent

\newpage
\section{Introduction}

Single transverse spin asymmetries (SSAs) in the strong interaction have a long history
since 70s and 80s when the large asymmetries were observed in 
$pp\to\Lambda^\uparrow X$\,\cite{Lambda} 
and $p^\uparrow p\to \pi X$\,\cite{E704} in the forward direction.
These data triggered lots of theoretical activities
to clarify the mechanism of the asymmetries\,\cite{Review}.  
Experimentally measurements of SSA at higher 
energies in $pp$ collisions have been performed at RHIC\,\cite{STAR,PHENIX,BRAHMS}, and
SSAs in semi-inclusive deep 
inelastic scattering (SIDIS) have been also reported\,\cite{hermes,compass}.
The SSA is a so-called ``naively T-odd'' observable proportional to
$(\pvec\times\qvec )\cdot\Svec_\perp$, and 
time-reversal invariance in QCD implies that
it only occurs from the interference between the
amplitudes which have different phases.  
In the literature, two QCD mechanisms have been used for describing the observed 
large SSAs.  One is based on the use of so-called ``T-odd'' distribution
and fragmentation functions which have explicit
intrinsic transverse momentum $\kvec_\perp$ of partons inside hadrons\,\cite{Sivers,Collins93}.
This mechanism describes the 
SSA as a leading twist effect in the region of small transverse momentum $p_T$  
of the produced hadron.
Factorization formula with the use of
$k_\perp$ dependent parton distribution functions
and fragmentation functions 
has been extended to SIDIS\,\cite{JMY05}, applying the method used for 
$e^+e^-$ annihilation\,\cite{CS81} and the Drell-Yan process\,\cite{CSS85}, and
the universality property
of those $k_\perp$-dependent functions
have been examined in great details\,
\cite{Collins02,BJY03,BMP03,CM04,BMP04}.  
Phenomenological applications of the ``T-odd'' functions 
have been also performed to interprete the existing data for SSAs\,\cite{Todd}. 

The other mechanism describes the SSA as a twist-3 effect 
in the collinear factorization, and is suited for describing
SSA in the large $p_T$ region\,\cite{ET82,QS91,QS99,KK00,KK01,Koike01,Koike03}.  
In this framework, SSA is connected 
to particular quark-gluon correlation functions on the lightcone. 
This method has been extensively applied to SSA in 
$pp$ collisions, such as direct photon production $p^\uparrow p\to \gamma X$\,\cite{QS91},
pion production 
$p^\uparrow p\to\pi X$\,\cite{QS99,KK00,Koike03}, 
and the hyperon polarization $pp\to\Lambda^\uparrow X$\,\cite{KK01} etc.
Although 
the above two mechanisms describe SSA in different
kinematic regions, it has been known that a soft-gluon pole function
appearing in the twist-3 mechanism is connected to a $\vec{k}_\perp$ moment of a
``T-odd'' function\,\cite{BMP03}.     
More recently, the authors of ref. \cite{JQVY06} showed that 
the two mechanisms give the identical 
SSA for the Drell-Yan process in the intermediate $p_T$ region,
unifying the two mechanisms.

In this paper, we study the SSA for large-$p_T$ pion production 
in SIDIS, $ep^\uparrow\to e\pi X$.~\footnote{The main result for SSA 
was reported in \cite{EK05} by two of the present authors (H.E. and Y.K.).}
The QCD factorization tells us that
the cross section for this process is generically 
expressed as convolution of three quantities, 
a distribution function for the transversely polarized nucleon,
a fragmentation function for the pion, and a partonic hard cross section. 
In the framework of the collinear factorization,
the SSAs at the leading order in QCD 
perturbation theory require participation of an additional, coherent gluon
into the partonic subprocess\,\cite{ET82}, and the corresponding gluon
field can be generated from either 
the nucleon or the pion, as shown in (a) and (b) of Fig. 1, respectively. 
These two contributions give rise to the 
following twist-3 cross section, denoting respectively as the (A) and (B) terms:
\beq
({\rm A}) & & G_F(x_1,x_2)\otimes D(z) \otimes \widehat{\sigma}_{\rm A},\nonumber\\
({\rm B}) & & \delta q(x) \otimes \widehat{E}_F(z_1,z_2) \otimes \widehat{\sigma}_{\rm B},
\label{twist3}
\eeq
where $G_F(x_1,x_2)$ and $\widehat{E}_F(z_1,z_2)$
are, respectively, 
the twist-3 distribution function for the transversely polarized nucleon and
the twist-3 fragmentation function for the pion; the definition of these twist-3 functions 
is given in (\ref{twist3distr}) and (\ref{twist3frag}) below. 
In (\ref{twist3}) there also appear the familiar twist-2 functions: 
$\delta q(x)$ is the quark transversity
distribution of the nucleon, and $D(z)$ is the usual 
quark/gluon fragmentation function
for the pion.
$\widehat{\sigma}_{\rm A,B}$ are the partonic hard cross sections, 
and we present their explicit formula in QCD perturbation theory.
We note that both distribution and fragmentation functions in the (A) term of (\ref{twist3})
are chiral-even, while both of those functions in the (B) term are chiral-odd.  
Accordingly, we have only one diagram in Fig.~1(b)
relevant to the (B) term.
We also note that there exist purely gluonic twist-3 distributions
for the transversely polarized nucleon which can contribute to
the (A) term in (\ref{twist3})\,\cite{Ji92}.  
The study of this term is beyond the scope of the present work,
and we shall restrict ourselves to the kinematic region where
the contribution from the quark-gluon correlation is expected to be dominant
compared with that from the three-gluon correlation.

It has been already shown that the twist-3 functions
$G_F$ and $\widehat{E}_F$ appearing in (\ref{twist3}) can potentially be important sources for
the 
large asymmetry $A_N$ observed at large $x_F$ for $p^\uparrow p\to \pi X$\,\cite{QS99,Koike03}.  
Therefore, our interest here
is to study how the above effects (A) and (B) contribute to the SSA in SIDIS,
if the overall
strengths of the relevant twist-3 
nonperturbative quantities $G_F(x_1,x_2)$, $\widehat{E}_F(z_1,z_2)$, 
as well as the transversity
$\delta q(x)$, 
are determined so as to 
reproduce
$A_N$ in $pp$ collisions\,\cite{E704,STAR}.   

The remainder of this paper is organized as follows:  In section 2, we present definition
of the twist-3 distribution and fragmentation functions
necessary in our analysis.  In section 3, after a brief description of the
kinematics for SIDIS, we present the polarized cross section formula
for the two contributions in (\ref{twist3}), focussing on the 
soft-gluon-pole contributions.  Azimuthal asymmetries from each
contribution are discussed in detail and their
numerical estimate 
is also presented.  A brief summary is given in section 5.  
Appendix provides an explicit relation between 
the different definitions
for 
the twist-3 distribution functions. 

\begin{figure}[t!]
\begin{center}
\epsfig{figure=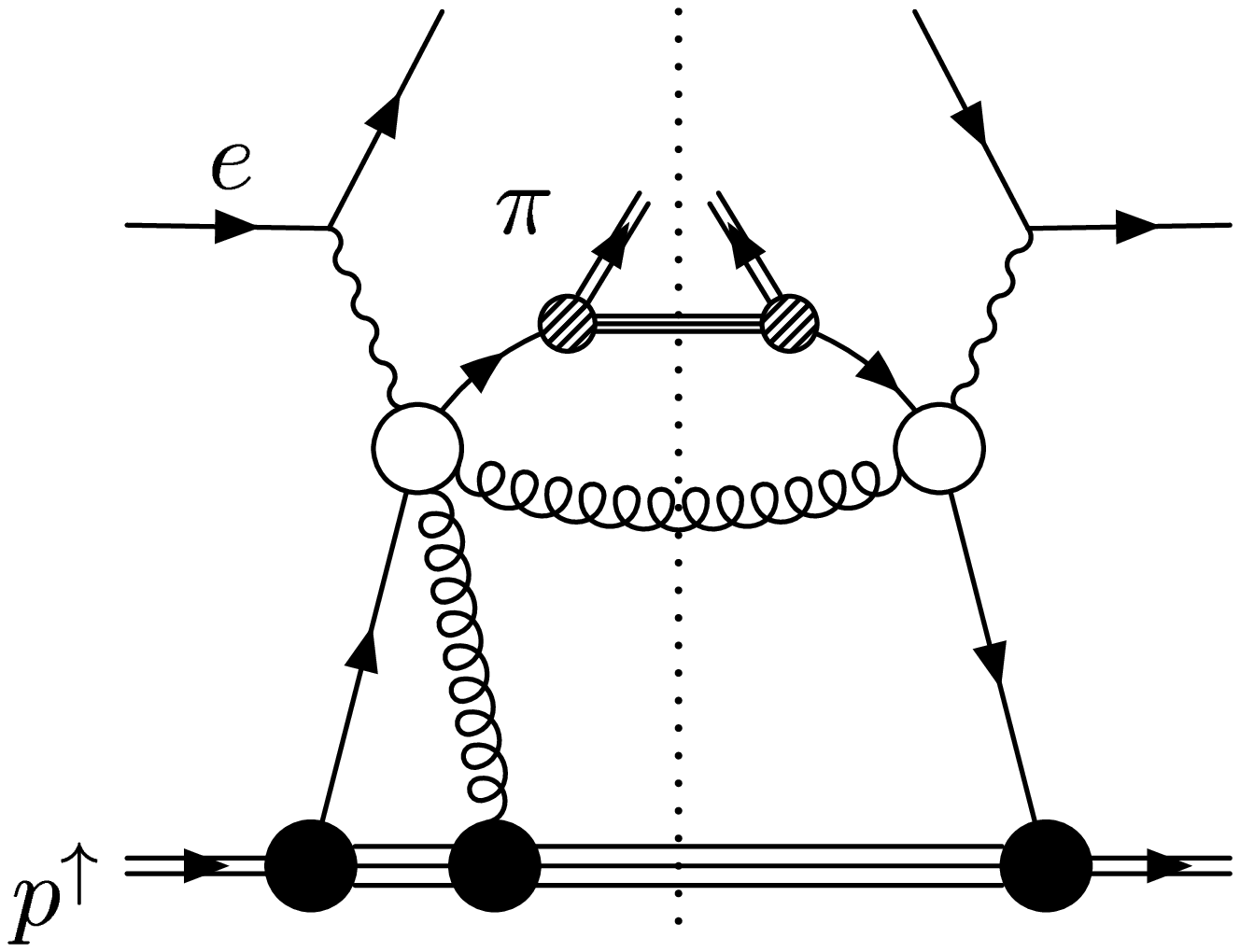,width=0.28\textwidth}
\hspace{5mm}
\epsfig{figure=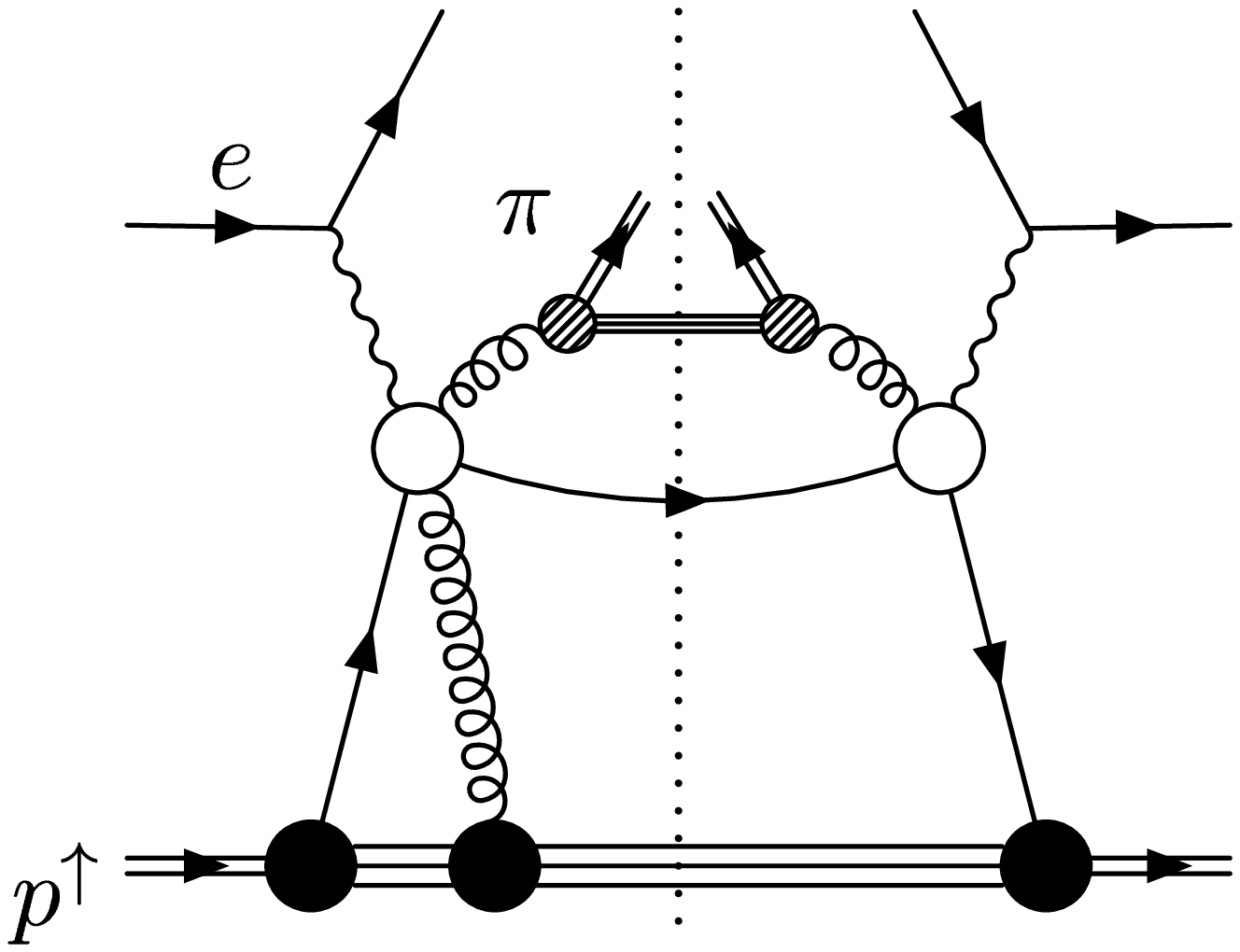,width=0.28\textwidth}
\hspace{10mm}
\epsfig{figure=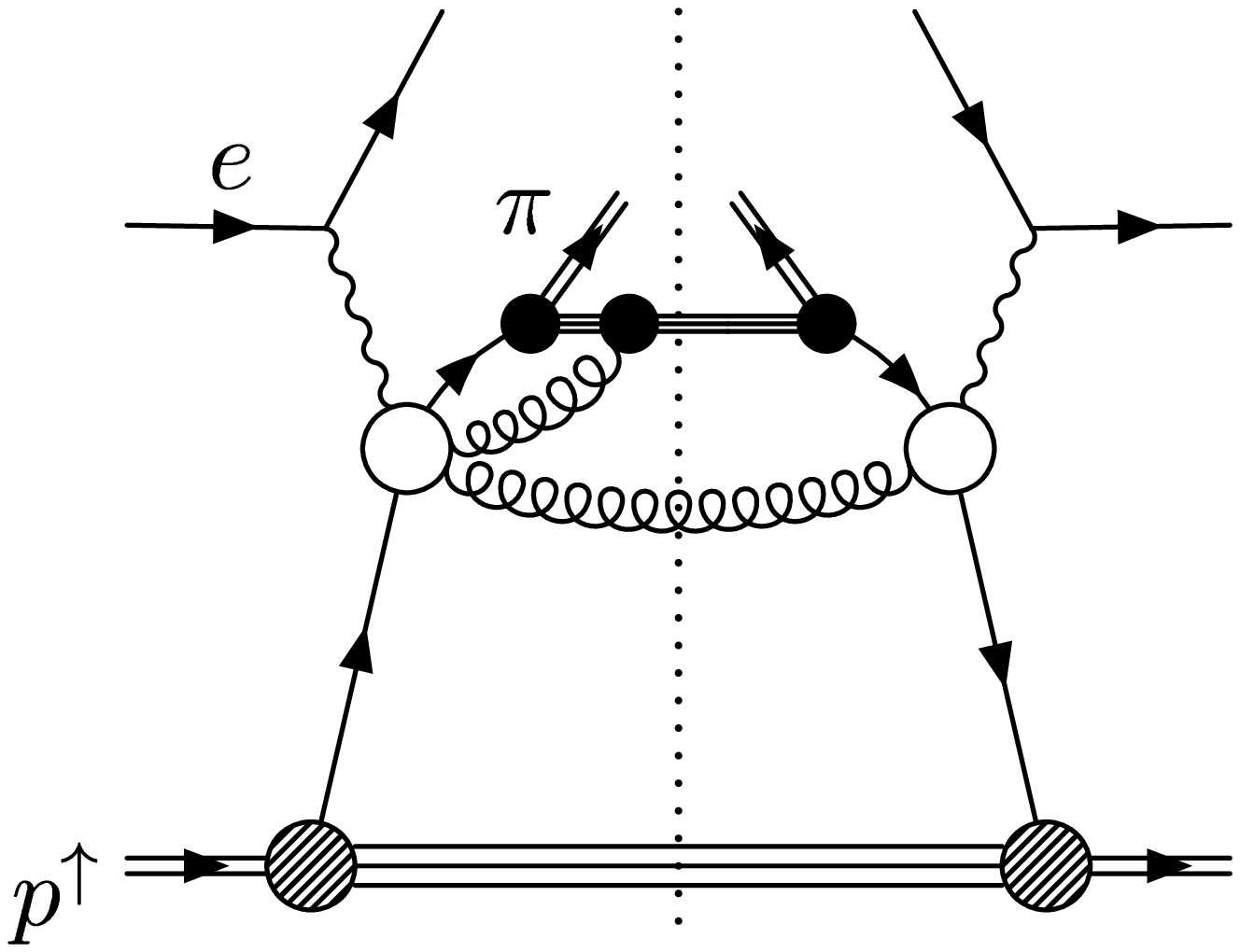,width=0.28\textwidth}
\end{center}

\hspace{5cm}(a)
\hspace{8cm}(b)

\caption{
(a) Contribution to $ep^\uparrow \to e\pi X$ from the twist-3
distribution $G_F$; in the left (right) diagram, a quark (gluon) fragments into $\pi +$anything.
(b) Contribution to $ep^\uparrow\to e\pi X$ from the
twist-3
fragmentation function $\widehat{E}_F$.
The white circles represent partonic hard scatterings.
Mirror diagrams obtained by the
interchange between left and right of the cut also contribute.
\label{fig1}}
\vspace*{0.cm}
\end{figure}

\section{Twist-3 quark-gluon correlation functions}

We first discuss the definition and basic properties 
of the twist-3 distributions for the transversely polarized nucleon,
which 
are relevant to 
$ep^\uparrow\to e\pi X$. As shown in (a) of Fig.~1, these distributions actually represent
the quark-gluon correlation inside the nucleon.
For the 
transversely polarized
nucleon with momentum $p^{\mu}$ and spin $S_\perp^\mu$, 
there are two independent twist-3 quark-gluon correlation functions which are defined as 
nucleon matrix element of nonlocal lightcone operator \,\cite{QS91,QS99,KT99} 
\begin{eqnarray}
& &\int {d\lambda\over 2\pi}\int{d\mu\over 2\pi}
e^{i\lambda x_1}e^{i\mu(x_2-x_1)}
\langle p\ S_\perp |\bar{\psi}_j(0)[0,\mu n]
{gF^{\alpha\beta}(\mu n)n_\beta}[\mu n, \lambda n]
\psi_i(\lambda n)|p\ S_\perp \rangle\nonumber\\
& &\qquad=
{M_N\over 4} \left(\pslash\right)_{ij} 
\epsilon^{\alpha pnS_\perp}
{G_F(x_1,x_2)}+ 
i{M_N\over 4} \left(\gamma_5\pslash\right)_{ij} 
S_\perp^\alpha
{\GFt(x_1,x_2)}+ \cdots,
\label{twist3distr}
\end{eqnarray}
where $\psi$ is the quark field, $F^{\alpha\beta}$ is the gluon field strength tensor,
$n^\mu$ is the light-like vector 
($n^2=0$) with $p\cdot n=1$,
$\epsilon^{\alpha pnS_\perp}=\epsilon^{\alpha\lambda\mu\nu}p_\lambda
n_\mu S_{\perp \nu}$ with $\epsilon_{0123}=1$, 
and the spin vector satisfies $S_\perp^2 = -1$, $S_\perp \cdot p=S_\perp \cdot n=0$.  
 $[\mu n,\lambda n]={\rm P} \exp\left[ ig \int_{\lambda}^{\mu}dt n\cdot A(tn) \right]$ 
represents the gauge-link which guarantees gauge invariance of the
nonlocal lightcone operator,
and 
``$\cdots$'' denotes twist-4 or higher-twist distributions. 
The correlation functions $G_F(x_1,x_2)$, $\GFt(x_1,x_2)$ 
are defined as dimensionless, and the nucleon mass $M_N$ in the RHS of (\ref{twist3distr})
represents a natural scale for chiral-symmetry breaking.  
Each correlation function depends on the two variables $x_1$ and $x_2$, 
where $x_1$ and $x_2 -x_1$ are the fractions of the lightcone momentum 
carried by the quark and the gluon, respectively, which are outgoing from the nucleon.
From P- and T-invariance in QCD, one can show the symmetry properties
\begin{equation}
G_F(x_1,x_2)=G_F(x_2,x_1),\qquad
\GFt(x_1,x_2)=-\GFt(x_2,x_1),
\label{sym}
\end{equation}
and from hermiticity both $G_F(x_1,x_2)$ and $\GFt (x_1,x_2)$ are real
functions.
These two correlation functions $G_F(x_1,x_2)$ and $\GFt (x_1,x_2)$ 
constitute a complete set of the 
twist-3 quark-gluon correlation functions 
for the transversely polarized nucleon (see e.g. ref.~\cite{KT99} for a concise discussion).

In the literature, however, another 
set of twist-3 quark-gluon correlation functions are also used\,\cite{QS91,QS99}:
Replacing the field strength
$gF^{\alpha\beta}(\mu n)n_\beta$ in the LHS in (\ref{twist3distr}) by the transverse components 
of the covariant derivative,
$D_{\perp}^\alpha(\mu n)=\partial_{\perp}^\alpha
-ig A_{\perp}^\alpha(\mu n)$, the same Lorentz decomposition in the RHS
defines the two twist-3 
distributions
$G_D(x_1,x_2)$ and $\GDt(x_1,x_2)$, in place of 
$G_F(x_1,x_2)$ and $\GFt(x_1,x_2)$, respectively. 
Physical meaning as well as the basic properties of 
$G_D(x_1,x_2)$ and $\GDt(x_1,x_2)$ are similar to $G_F(x_1,x_2)$ and $\GFt(x_1,x_2)$, except that 
P- and T-invariance implies $G_D(x_1,x_2)=-G_D(x_2,x_1)$, $\GDt(x_1,x_2)=\GDt(x_2,x_1)$.

Overcomplete set of four 
twist-3 correlation 
functions $G_F(x_1,x_2)$, $\GFt(x_1,x_2)$, $G_D(x_1,x_2)$, and $\GDt(x_1,x_2)$ 
for transversely polarized nucleon
has been conveniently used on a case-by-case 
basis to express the cross section formula in the literature.   
Here we present an explicit relation between the 
``$F$-type'' distributions ($G_F(x_1,x_2), \GFt(x_1,x_2)$) 
and the ``$D$-type'' functions ($G_D(x_1,x_2), \GDt(x_1,x_2)$).
For this purpose, we refer to  
the operator identity with manifest gauge invariance, 
\beq
D^\alpha(\mu n)[\mu n,\lambda n]=
i\int_{\lambda}^{\mu}dt [\mu n,t  n]gF^{\alpha \beta}(tn)n_{\beta} [t n,\lambda n]+ 
[\mu n,\lambda n]D^\alpha(\lambda n),   
\label{id}
\eeq
which relates (\ref{twist3distr}) 
with the corresponding matrix element for the $D$-type distributions.
One then obtains
\beq
G_D(x_1,x_2)&=&P{1\over x_1-x_2}G_F(x_1,x_2),
\label{FDvector}\\
\GDt(x_1,x_2)&=&\delta(x_1-x_2)\gt(x_1)+
P{1\over x_1-x_2}\GFt(x_1,x_2), 
\label{FDrelation}
\eeq
where $P$ denotes the principal value, and 
\beq
\lefteqn{\gt(x)=\frac{-i}{M_{N}}\int{d\lambda\over 2\pi}e^{i\lambda x}\la p\ S_\perp |
\bar{\psi}(0) [0 ,\lambda n] \gamma_5 \nslash S_\perp\cdot D(\lambda n)
\psi(\lambda n)|p\ S_\perp \ra}\nonumber\\
+&&
\!\!\!\!
{1\over 2M_N }\int{d\lambda\over 2\pi}e^{i\lambda x}\int_{-\infty}^\infty d\mu\, 
\epsilon(\mu-\lambda)\,\la p\ S_\perp |
\bar{\psi}(0) [0 ,\lambda n]\gamma_5 \nslash g
F^{S_\perp n}(\mu n)[\mu n,\lambda n]
\psi(\lambda n)|p\ S_\perp \ra,\nonumber\\
\label{gt}
\eeq
with $\epsilon(\mu-\lambda)=2\theta(\mu-\lambda) -1$.
In the RHS of (\ref{FDvector}), the term proportional to
$\delta(x_1-x_2)$
vanishes due to the anti-symmetry of $G_D(x_1,x_2)$ as stated
above.   
In Appendix, we demonstrate that $\gt(x)$ can be reexpressed in terms of
the $F$-type distributions $G_F(x_1,x_2)$ and $\GFt(x_1,x_2)$, 
and the twist-2 quark helicity distribution $\Delta q(x)$ of the nucleon.
The relations (\ref{FDvector}) and (\ref{FDrelation}) 
show that the $D$-type functions are more singular than
the $F$-type functions at the soft gluon point $x_1=x_2$, while
they are proportional to each other for $x_1\neq x_2$.  
In our analysis for $ep^\uparrow\to e\pi X$ below, we assume
that there is no singularity in 
the $F$-type functions, in particular, 
that $G_F(x_1,x_2)$ is finite at $x_1=x_2$ ($\GFt(x_1,x_1)=0$, due to (\ref{sym})).  

Similarly to the distributions, 
one can also construct the twist-3 
fragmentation function for the pion with momentum $\ell$ as\,\cite{Koike01,BMP03,Koike03}
\begin{eqnarray}
& &{1\over N_c}\sum_X
\int {d\lambda\over 2\pi}\int{d\mu\over 2\pi}
e^{-i{\lambda\over z_1}}e^{-i\mu({1\over z_2}-{1\over z_1})}
\langle 0|[-\infty w,0]\psi_i(0)|\pi(\ell)X\rangle \nonumber\\
& & \qquad\qquad\qquad\times
\langle \pi(\ell)X|
\bar{\psi}_j(\lambda w)[\lambda w,\mu w]{gF^{\alpha\beta}(\mu w)w_\beta}
[\mu w,-\infty w]|0\rangle\nonumber\\
& &\qquad=
{
{M_N}\over 2z_2}
\left(\gamma_5\ellslash\gamma_\nu\right)_{ij}
\epsilon^{\nu \alpha w\ell}{
\widehat{E}_F(z_1,z_2)}+\cdots,
\label{twist3frag}
\end{eqnarray}
where we introduced the light-like vector $w$ by the relation $\ell\cdot
w=1$, and 
``$\cdots$'' denotes twist-4 or higher, and we again use the nucleon mass $M_N$
as a generic QCD mass scale
in order to define $\widehat{E}_F(z_1,z_2)$ as dimensionless.
A fragmentation function obtained by shifting the field strength
${gF^{\alpha\beta}(\mu n)w_\beta}$ from the matrix element
$\langle \pi(\ell)X|\cdots|0\rangle$ to
$\langle 0|\cdots|\pi(\ell)X\rangle$ is expressed in terms of 
$\widehat{E}_F^*(z_2,z_1)$.  
Unlike (\ref{sym}) for the twist-3 
distributions, time reversal invariance does not bring any constraint
on the symmetry properties of the twist-3 fragmentation functions.  
As in the case for the twist-3 distribution, we can also
define another 
twist-3 quark-gluon fragmentation function $\widehat{E}_D(z_1,z_2)$, 
replacing ${gF^{\alpha\beta}(\mu w)w_\beta}$ by the covariant derivative
$\overleftarrow{D}_\perp^\alpha (\mu w)$ in (\ref{twist3frag})\,\cite{Ji94}.  
The relation between $\widehat{E}_D(z_1,z_2)$
and $\widehat{E}_F(z_1,z_2)$
can be easily obtained by using 
the identity (\ref{id}) as 
\beq
\widehat{E}_D(z_1,z_2)=\delta\left({1\over z_1}-{1\over
z_2}\right)\widetilde{e}(z_1)
+P{1\over (1/z_1)-(1/z_2)}\widehat{E}_F(z_1,z_2),
\eeq
where
\beq
\widetilde{e}(z)&=&
{z\over 4M_N}{1\over N_c}
\sum_X\int{d\lambda\over 2\pi}\,e^{-i\lambda/z}
\la 0|[-\infty w,0]\gamma^\nu\wslash\gamma_5\psi(0)|\pi(\ell)X\ra\nonumber\\
& &\qquad \times
\la\pi(\ell)X|\bar{\psi}(\lambda w)\overleftarrow{D}^\alpha(\lambda w)
[\lambda w,-\infty w]|0\ra \epsilon_{\nu\alpha}^{\ \ \,w\ell}
\nonumber\\
& &
-{iz\over 8M_N}{1\over N_c}\sum_X
\int{d\lambda\over 2\pi}\,e^{-i\lambda/z}\int_{-\infty}^\infty d\mu\,\epsilon(\mu-\lambda)
\la 0|[-\infty w,0]\gamma^\nu\wslash\gamma_5\psi(0)|\pi(\ell)X\ra\nonumber\\
& &\qquad \times
\la\pi(\ell)X|\bar{\psi}(\lambda w)[\lambda w,\mu w]g{F}^{\alpha w}(\mu w)
[\mu w,-\infty w]|0\ra \epsilon_{\nu\alpha}^{\ \ \,w\ell}.  
\eeq
Here we emphasize again that $\widehat{E}_D(z_1,z_2)$ is more singular
compared to $\widehat{E}_F(z_1,z_2)$ at $z_1=z_2$, while they are proportional to each other
for $z_1\neq z_2$.   
As will be discussed in the next section, the
cross section for $ep^\uparrow\to e\pi X$ receives important contribution from the ``soft gluon point'',  
$x_1=x_2$ or $z_1=z_2$, of the twist-3 functions defined above. We thus use
the corresponding $F$-type functions
to express those contributions in the cross section.
Other terms which receive contributions with $x_1\neq x_2$, $z_1\neq z_2$ are 
expressed in terms of either $D$-type or $F$-type functions without any subtlety.

\section{Single spin asymmetry for $ep^\uparrow\to e\pi X$}

\subsection{Kinematics}

Here we present a brief description of the kinematics for the SIDIS, 
$e(k)+p(p, S_{\perp})\to e(k')+\pi(\ell)+X$. (See refs.~\cite{MOS92,KN03} for the detail.)  
We have five independent Lorentz invariants, 
$S_{ep}, x_{bj},Q^2,z_f$, and $q_T^2$.  
The center of mass energy squared, $S_{ep}$, for the initial
electron and the proton is
\beq
S_{ep}=(p + k)^2 \simeq 2p\cdot k \; ,
\eeq
ignoring masses. The conventional DIS variables are defined 
in terms of the virtual photon momentum $q=k-k'$ as
\beq
x_{bj}={Q^2\over 2p\cdot q} \; , \qquad Q^2 =-q^2=-(k-k')^2. 
\label{xbj}
\eeq
For the final-state pion, we introduce the scaling variable
\beq
z_f={p\cdot \ell\over p\cdot q} \; .
\eeq
Finally, we define the ``transverse'' component of $q$, which is orthogonal to
both $p$ and $\ell$:
\beq
q_t^\mu=q^\mu- {\ell\cdot q\over p\cdot \ell}p^\mu -
{p\cdot q\over p\cdot \ell}\ell^\mu \; .
\eeq
$q_t$ is a space-like vector, and we denote its magnitude by
\beq
q_T = \sqrt{-q_t^2} \; .
\eeq
To completely specify the kinematics, we need to choose a reference
frame. We shall work in the so-called {\it hadron frame}~\cite{MOS92},
which is the Breit frame of the virtual photon and the initial proton:
\beq
q^\mu &=& (0,0,0,-Q) \; ,\\
p^\mu &=& \left( {Q\over 2x_{bj}},0,0,{Q\over 2x_{bj}}\right) \; .
\eeq
Further, in this frame the outgoing pion is taken to be in the $xz$ plane:
\beq
\ell^\mu = {z_f Q \over 2}\left( 1 + {q_T^2\over Q^2},{2 q_T\over Q},0,
{q_T^2\over Q^2}-1\right) \; .
\label{eq2.p_B}
\eeq
As one can see, the transverse momentum of the pion is in this 
frame given by $\ell_{T}=z_f q_T$. This is true for any frame in which the
3-momenta of the virtual photon and the initial proton are collinear.
By introducing the angle $\phi$ between the hadron plane and the lepton plane, 
the lepton momentum can be parameterized as
\beq
k^\mu={Q\over 2}\left( \cosh\psi,\sinh\psi\cos\phi,
\sinh\psi\sin\phi,-1\right) \; ,
\label{eq2.lepton}
\eeq
and one finds
\beq
\cosh\psi = {2x_{bj}S_{ep}\over Q^2} -1 \; .
\label{eq2.cosh}
\eeq
We parameterize the transverse spin vector of the initial proton $S_\perp^\mu$
as
\beq
S^\mu_\perp = (0,\cos\Phi_S,\sin\Phi_S,0),
\label{phis}
\eeq
where $\Phi_S$ represents the azimuthal angle of $\vec{S}_\perp$ measured from
the hadron plane.
With the above definition, 
the cross section for $ep^\uparrow\to e\pi X$ can be expressed in terms of
$S_{ep}$, $x_{bj}$, $Q^2$, $z_f$, $q_T^2$, $\phi$ and $\Phi_S$ in the hadron
frame.  Note that $\phi$ and $\Phi_S$ are invariant under boosts in the 
$\vec{q}$-direction, so that the cross section presented below is
the same in any frame where $\vec{q}$ and $\vec{p}$ are collinear.  

\subsection{Twist-3 polarized cross section}

We now proceed to derive the cross section for $ep^\uparrow\to e\pi X$,
applying the method developed 
in \cite{QS91,QS99}.  We use Feynman gauge in the actual calculation.  
In this framework, the phase necessary for SSAs is provided from 
the partonic hard cross section.  To illustrate, consider the diagrams
in the first line of 
Fig.~2(a),
which are some of the leading order 
contributions in perturbation theory corresponding to Fig.~1(a).
These diagrams involve the 
coherent gluon from the nucleon, participating in the partonic hard scattering.
After the collinear expansion 
in the transverse momenta $k_\perp$ of the partons inside nucleon to $O(k_{\perp})$, 
the ``external'' partons for the partonic hard cross section are collinear to their parent hadron. 
Denoting the momenta of the external quark lines
on the LHS of the cut as $x_1 p$ 
and $\ell/z$, and the momentum of the coherent gluon as $(x_2 -x_1) p$, 
where $x_1 , x_2$, and $z$ are the corresponding momentum fractions to be integrated over,
the parton propagator coupled to the coherent gluon 
gives a factor $1/(x_1 - x_2 + i\varepsilon )$.  
Then the integration for the parton momentum fraction produces the phase   
from the pole contribution at the soft gluon point 
$x_1 = x_2$ (soft-gluon-pole (SGP))\,\cite{QS91}, which may be evaluated
through the distribution identity
\beq
{1\over x_1 -x_2 + i\varepsilon } = P{1\over x_1-x_2} - i\pi \delta(x_1 - x_2).
\label{distribution}
\eeq
It is straightforward to see that 
another propagator in the LHS of the cut in the first and third
diagrams in the first line in Fig.~2(a) 
is proportional to $1/( x_1 - i\varepsilon ) = P(1/ x_1) + i\pi
\delta(x_1)$,
which produces 
the soft-fermion-pole (SFP)
contribution at $x_1=0$\,\cite{QS91}.  
Similarly another propagator in the LHS of the cut in the second and fourth
diagrams in the first line in Fig.~2(a)
is proportional to $1/( x_1 -x_{bj} +i\varepsilon ) = P\{1/ (x_1-x_{bj})\} - i\pi
\delta(x_1-x_{bj})$,
which produces 
the hard-pole (HP)
contribution at $x_1=x_{bj}$\,\cite{Guo98}.   
There are other leading order diagrams of the type of Fig.~1(a),
which receive the SFP and HP contributions. 
As anticipated, the SGP contributions, 
as well as the SFP and HP contributions, are eventually associated 
with the quark-gluon correlation beyond twist-2,
and are expressed by a complete set of the 
twist-3 quark-gluon correlation functions, $G_F(x_1,x_2)$ and $\GFt (x_1,x_2)$ of 
(\ref{twist3distr}).  
From the symmetry (\ref{sym}) under $x_1 \leftrightarrow x_2$, 
$\GFt(x_1,x_2)$ does not contribute to the SGP at $x_1=x_2$.
On the other hand, as a result of the collinear expansion, some of the
SGP contributions appear with the derivative like ${d\over dx}G_F(x,x)$,
while SFP
and HP contributions
do not appear with such derivative.\footnote{
The SFP and HP contributions can be straightforwardly
calculated in the lightcone gauge $A^+=0$,
in which we can make direct translation $A^\perp\to{iF^{\perp n}/(x_1-x_2)}$, corresponding to the 
relations (\ref{FDvector}), (\ref{FDrelation}) for $x_1 \neq x_2$,
and no collinear expansion is necessary.  Thus no derivative terms appear
for those contributions.  This feature was also confirmed by explicit
calculation in the Feynman gauge in a recent paper \cite{JQVY06SIDIS}.}  
In the large $x$ region, one has $|x{d\over dx}G_F(x,x)| \gg |G_F(x,x)|$, 
since $G_F(x,x)$ is expected to behave as $G_F(x,x)\sim (1-x)^\beta$ ($\beta >0$).  
Accordingly, it is justified to keep only the derivative terms with ${d\over dx}G_F(x,x)$ 
in the SGP contributions
for the case
where only large $x$ region 
is relevant.

Similarly, the diagrams in Fig.~2(b),
which are some of the leading order 
contributions in perturbation theory corresponding to Fig.~1(b),
also receive the pole
contributions.
After the collinear expansion, the quark momenta
entering and leaving
the pion fragmentation function can be parameterized as $\ell/z_1$ and
$\ell/z_2$, respectively.
In this case, the pole contributions, which give the phase for SSA, come from
the 
SGP contribution at $z_1=z_2$ and the HP contribution
at $z_1=z_f (1-q_T^2 /Q^2 )$.
But there appears no SFP
contribution, since $1/z_i\neq 0$ ($i=1,\ 2$).  
Again, some of the SGP contributions appear with the
derivative of $\widehat{E}_F$, while no derivative appears
for the HP contributions.~\footnote{In principle, 
the principal-value part of the distribution identity of the type (\ref{distribution}),
combined with the imaginary part of $\widehat{E}_F$, could produce SSA. 
Discussion on such effects
associated with nonperturbative generation of 
the strong interaction phase is beyond the scope of this work.
Correspondingly, in the following, we shall assume
$\widehat{E}_F(z_1,z_2)$ is real.}

In our present study, we shall focus on the derivative terms of the SGP contributions
following the previous studies\,\cite{QS99,KK00,KK01,Koike03}.
This approximation should be valid if either $x_{bj}$ or $z_f$ is large, where
the main contribution is from large-$x$ and large-$z$ 
regions (see the cross section formula derived below). 
Also, physically, one expects that the SFP and HP contributions are suppressed compared to 
the SGP contributions, because there may be lots of soft gluons in the hadrons; in particular,
for the SFP contributions, not much soft fermions are expected to exist in a hadron.  

\begin{figure}[t!]
\begin{center}
\epsfig{figure=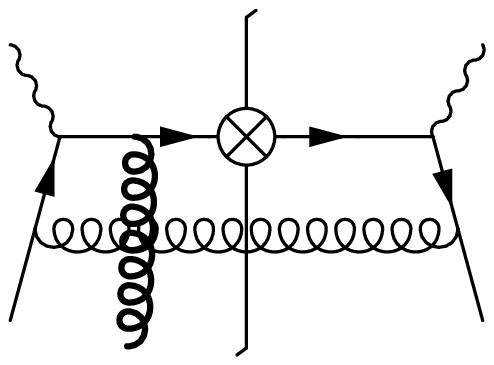,width=0.2\textwidth}
\hspace{2mm}
\epsfig{figure=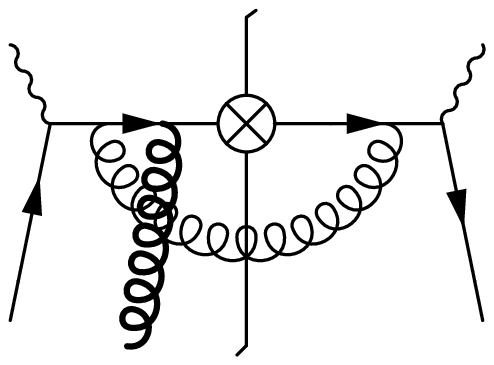,width=0.2\textwidth}
\hspace{2mm}
\epsfig{figure=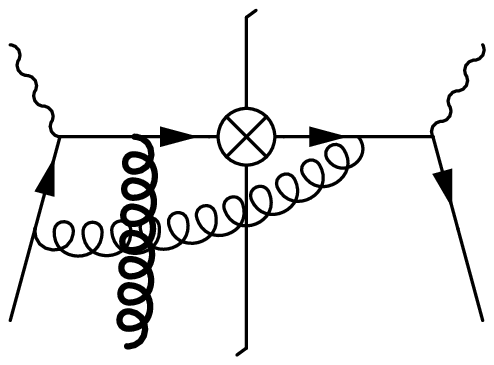,width=0.2\textwidth}
\hspace{2mm}
\epsfig{figure=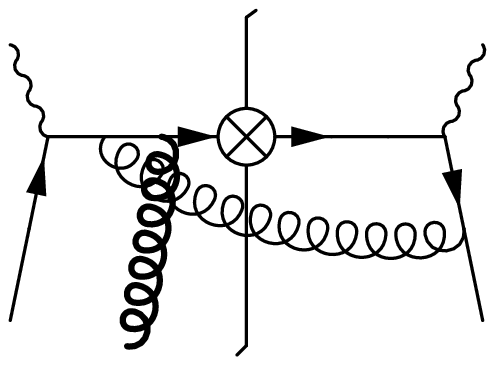,width=0.2\textwidth}
\hspace{2mm}
\epsfig{figure=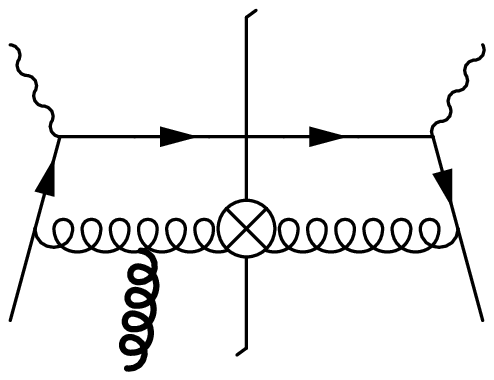,width=0.2\textwidth}
\hspace{2mm}
\epsfig{figure=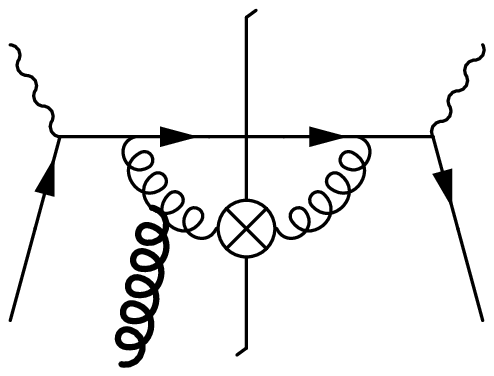,width=0.2\textwidth}
\hspace{2mm}
\epsfig{figure=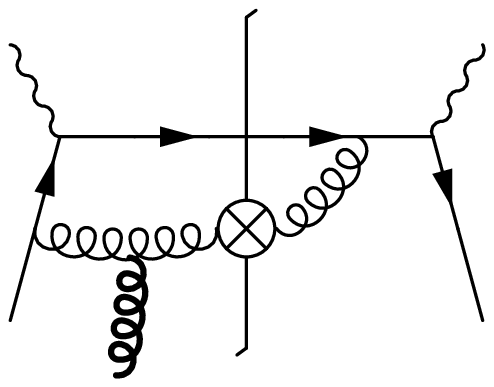,width=0.2\textwidth}
\hspace{2mm}
\epsfig{figure=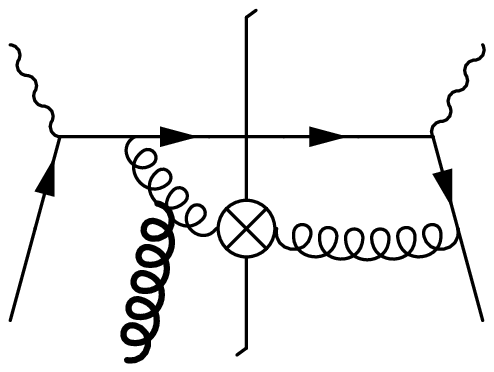,width=0.2\textwidth}

(a)

\vspace{0.5cm}

\epsfig{figure=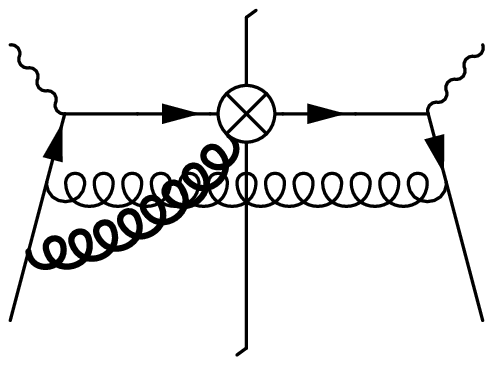,width=0.2\textwidth}
\hspace{2mm}
\epsfig{figure=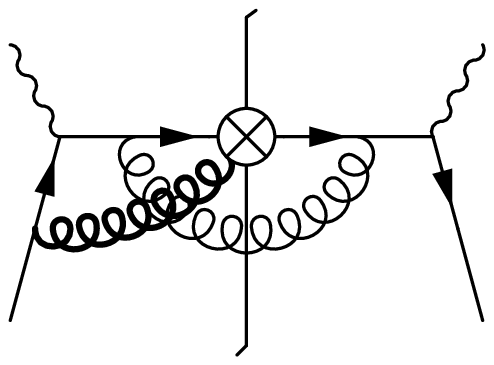,width=0.2\textwidth}
\hspace{2mm}
\epsfig{figure=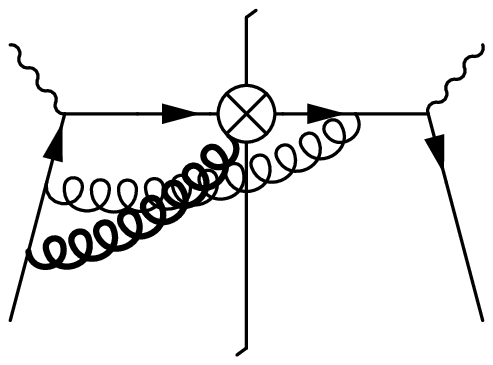,width=0.2\textwidth}
\hspace{2mm}
\epsfig{figure=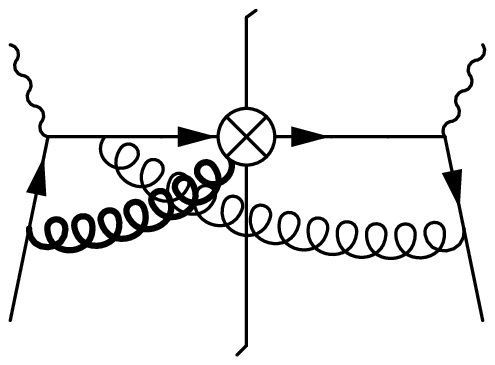,width=0.2\textwidth}

(b)
\end{center}
\caption{
(a) and (b) represent Feynman diagrams which give rise to SGP
 contributions
for $G_F$ and $\widehat{E}_F$, respectively.  $\otimes$ denotes
 fragmentation insertions.  In both cases, mirror diagrams also
 contribute. 
\label{fig2}}
\vspace*{0.cm}
\end{figure}

The diagrams which receive the SGP contributions of the type (A) in (\ref{twist3})
are shown in Fig. 2(a).  
By including only the derivative terms of the SGP contribution, 
one obtains the spin-dependent differential cross section as  
\beq
{d^5\sigma^{({\rm A})}\over dx_{bj} dQ^2 dz_f dq_T^2 d\phi}
&=& {\alpha_{em}^2 \alpha_s \over 8\pi x_{bj}^2 
S_{ep}^2 Q^2}\left( {M_N q_T\pi\over Q^2}\right)x_{bj}^2
\sin\Phi_S
\sum_a e_a^2
\int_{x_{min}}^1\,{dx\over x}\int_{z_{min}}^1\,{dz\over z}\,
\nonumber\\
&\times & \left( {d\over dx}{G}_F^a(x,x)\right)
\left( D_a(z)C_a\widehat{\sigma}_{qq}^{\rm A} + D_g(z)C_g\widehat{\sigma}_{gq}^{\rm A} \right)
{\zhat\over (1-\zhat)^2}
\delta(x-x_0),\nonumber\\ 
\label{crossA}
\eeq
where
$\alpha_{em}=e^2/4\pi$ is the QED coupling constant, and 
we have introduced the variables
\beq
\xhat={x_{bj}\over x}\; ,\qquad \zhat={z_f\over z}\; ,
\eeq
\beq
x_{min}=x_{bj}\left( 1 + {z_f\over 1-z_f}{q_T^2\over Q^2}\right)\; ,\qquad
z_{min}=z_{f}\left( 1 + {x_{bj}\over 1-x_{bj}}{q_T^2\over Q^2}\right) \; ,
\label{eq3.2}
\eeq
\beq
x_0=x_{bj}\left( 1 + {\zhat\over 1-\zhat}{q_T^2\over Q^2}\right). 
\eeq
In (\ref{crossA}), 
$G_F^a$ denotes
the twist-3 distribution for the quark flavor $a$ defined in (\ref{twist3distr}), 
$D_a(z)/D_g(z)$ is the unpolarized 
quark (flavor $a$)/gluon fragmentation function for the pion, and
$e_a$ is the electric charge for the quark flavor $a$. 
The color factor $C_{a,g}$ is defined as
$C_q=C_F-N_c/2= -1/(2N_c)$ 
for quark and $C_{\bar{q}}=1/(2N_c)$ for anti-quark, and $C_g=N_c/2$ for gluon.  
The partonic hard cross section $\widehat{\sigma}_{jq}^{\rm A}$ ($j=q,\ g$)
in (\ref{crossA}) is written as
\beq
\widehat{\sigma}_{jq}^{\rm A} =\left(1+\cosh^2\psi\right)\,\widehat{\sigma}^1_{jq} -2\widehat{\sigma}_{jq}^2
-\sinh 2\psi\cos\phi\,\widehat{\sigma}^3_{jq} + \sinh^2\psi\cos 2\phi\,\widehat{\sigma}^4_{jq}, 
\qquad (j=q,g)
\eeq
where
\beq
\widehat{\sigma}^1_{qq}&=&2\xhat\zhat\left\{{1\over Q^2q_T^2}
\left({Q^4\over \xhat^2\zhat^2} + \left(Q^2-q_T^2\right)^2\right) +6\right\} \; ,
\nonumber\\
\widehat{\sigma}^2_{qq}&=&2\widehat{\sigma}^4_{qq}=8\xhat\zhat \; ,\nonumber\\
\widehat{\sigma}^3_{qq}&=&4\xhat\zhat{1\over Qq_T}(Q^2+q_T^2) \; ,
\label{evenqq}
\eeq
and 
\beq
\widehat{\sigma}^1_{gq}&=&2\xhat(1-\zhat)\left\{{1\over Q^2q_T^2}
\left({Q^4\over \xhat^2\zhat^2} + {(1-\zhat)^2\over \zhat^2}
\left(Q^2-{\zhat^2q_T^2\over (1-\zhat)^2}
\right)^2\right) +6\right\} \; ,
\nonumber\\
\widehat{\sigma}^2_{gq}&=&
2\widehat{\sigma}^4_{gq}=8\xhat(1-\zhat) \; ,\nonumber\\
\widehat{\sigma}^3_{gq}&=&-4\xhat(1-\zhat)^2{1\over \zhat Qq_T}
\left\{Q^2+
{\zhat^2 q_T^2\over (1-\zhat)^2}\right\} \; .
\label{evenqg}
\eeq
We note that the above $\widehat{\sigma}_{jq}^k$ ($j=q,\,g;\ k=1,\cdots,4$)
for the twist-3 cross section (\ref{crossA})
are the same as the corresponding unpolarized hard cross
sections\,\cite{Mendez78,KN03}.  
The suppression factor $(M_Nq_T/Q^2)$ in (\ref{crossA}) characterizes twist-3
nature of the cross section.  
We also note that, in the present 
case where either $x_{bj}$ or $z_f$ is large, the main contribution 
to (\ref{crossA}) is from regions with large $x$ and $z$, and this allows us to 
include only the contributions from the valence component
of the distribution functions and the favored component for the fragmentation functions.

Likewise 
we focus on 
the SGP contributions for the (B) term in (\ref{twist3}), which come from the diagrams
shown in Fig. 2(b).
We keep only the corresponding derivative terms of $\widehat{E}_F$, 
following \,\cite{QS99,KK00,KK01,Koike03}.
The spin-dependent cross section is obtained to be  
\beq
{d^5\sigma^{({\rm B})}\over dx_{bj} dQ^2 dz_f dq_T^2 d\phi}
&=& {\alpha_{em}^2 \alpha_s \over 8\pi x_{bj}^2 S_{ep}^2 Q^2}
\left( {M_N \pi\over Q}\right)\sum_a e_a^2
\int_{x_{min}}^1\,{dx\over x}\int_{z_{min}}^1\,{dz\over z}\,
\delta q_a(x) 
\nonumber\\
& &  \times 
\left\{
2\left[-z^2{\partial \over \partial z_1}\widehat{E}_F^a(z_1,z)\right]_{z_1=z}
+ {(1-2\xhat)Q^2\over\xhat q_T^2 + (1-\xhat)Q^2}
\left(-z^2{d\over dz}\widehat{E}_F^a(z,z)\right)
\right\}\nonumber\\
& &\times C_a\widehat{\sigma}^{\rm B} {\xhat^2\over
1-\xhat}\delta(z-z_0),   
\label{crossB}
\eeq
where $\widehat{E}_F^a$ is the twist-3 fragmentation function for the
quark flavor $a$ defined in (\ref{twist3frag}), 
the color factor $C_a$ is defined as $C_q=-1/(2N_c)$ and $C_{\bar{q}}=1/(2N_c)$
for
quark and anti-quark, respectively,  
the variable $z_0$ is given by
\beq
z_0=z_{f}\left( 1 + {\xhat\over 1-\xhat}{q_T^2\over Q^2}\right) \;,
\eeq
and the partonic hard cross section $\widehat{\sigma}^{\rm B}$ is defined as 
\beq
\widehat{\sigma}^{\rm B} &=& 
(1+\cosh^2\psi)\sin\Phi_S\left( {-4Qq_T\over \xhat(Q^2+q_T^2)} \right)
+ \sinh 2\psi\sin(\Phi_S-\phi)\left({4Q^2\over \xhat (Q^2 + q_T^2)}\right)\nonumber\\
& &\qquad+ \sinh^2\psi\sin(\Phi_S-2\phi)\left( {-4Q^3\over \xhat q_T (Q^2 + q_T^2 )}\right).  
\eeq
Here again we note the presence of the
suppression factor $M_N/Q$ in (\ref{crossB}) which characterizes the twist-3 cross section.  
Similarly to (\ref{crossA}), with
either $x_{bj}$ or $z_f$ being large, we can restrict the sum over $a$
to the contributions from the valence component
of the distribution functions and the 
favored component for the fragmentation functions in (\ref{crossB}).
Note that,
in (\ref{crossB}), there are two terms with 
$\left[(\partial/\partial z_1)\widehat{E}_F(z_1,z)\right]_{z_1=z}$ and
$(d/ d z)\widehat{E}_F(z,z)$,  
since $\widehat{E}_F(z_1,z_2)$ does not 
have definite symmetry property under $z_1\leftrightarrow z_2$.

\subsection{Azimuthal asymmetry}

A remarkable feature of the polarized cross section for 
$ep^\uparrow\to e\pi X$ is its characteristic 
azimuthal dependence.  
From (\ref{crossA}) and (\ref{crossB}), 
one finds the cross sections can be decomposed as
\beq
{d^5\sigma^{({\rm A})}\over dx_{bj} dQ^2 dz_f dq_T^2 d\phi}
=\sin\Phi_S\left( \sigma_0^{\rm A} 
+\sigma_1^{\rm A} \cos(\phi) +\sigma_2^{\rm A}\cos(2\phi)\right), 
\label{azimuthA}
\eeq
and
\beq
{d^5\sigma^{({\rm B})}\over dx_{bj} dQ^2 dz_f dq_T^2 d\phi}
= \sigma_0^{\rm B}\sin(\Phi_S) 
+\sigma_1^{\rm B} \sin(\Phi_S-\phi) +\sigma_2^{\rm B}\sin(\Phi_S-2\phi) .
\label{azimuthB}
\eeq
These should be compared with the similar decomposition of the twist-2 unpolarized
cross section\,\cite{Mendez78,MOS92}: 
\beq
{d^5\sigma^{\rm unpol}\over dx_{bj} dQ^2 dz_f dq_T^2 d\phi}
=\sigma_0^O +\sigma_1^O \cos(\phi) +\sigma_2^O\cos(2\phi).  
\label{azimuthunpol}
\eeq
As in (\ref{eq2.lepton}) and (\ref{phis}),
our azimuthal angles $\phi$ and $\Phi_S$ are defined in terms
of the hadron plane.   If one uses the {\it lepton plane} as a reference plane to
define the azimuthal angle of the spin vector
of the initial proton as $\phi_S$, and that of the hadron plane as $\phi_h$,
as employed in \cite{hermes}, one has the relation
$\Phi_S=\phi_h-\phi_S$ and $\phi=\phi_h$.  From this relation,
one sees that azimuthal dependence of 
the $\sigma_0^{\rm A}$ term in (\ref{azimuthA}) is the same as the Sivers
effect ($\propto
\sin(\phi_h-\phi_S)$), and that of $\sigma_2^{\rm B}$ term in (\ref{azimuthB}) is the same as
the Collins effect ($\propto \sin(\phi_h+\phi_S)$).  
Our cross sections (\ref{azimuthA}), (\ref{azimuthB}) have additional azimuthal components 
which are absent in the leading order cross section formula
obtained with Sivers and Collins functions.

We now proceed to estimate the SSA 
in SIDIS by using the obtained formulae (\ref{crossA}) and (\ref{crossB}). 
So far we don't have definite information on the 
nonperturbative functions, the twist-3 correlations $G_F^a$, $\widehat{E}_F^a$ 
and the transversity
distribution $\delta q_a$.\footnote{
$\widehat{E}_F^a(z,z)$ is related to the $k_\perp^2$-moment of the
difference of the two types of the Collins fragmentation functions with the ``future-pointing''
and ``past-pointing'' gauge links \cite{BMP03}. 
Although it's been shown in \cite{CM04} that one can derive factorization
for SIDIS as well as $e^+e^-$ annihilation by using the 
``future-pointing" gauge link, no definite relation is known between the two types of the Collins functions
with different gauge links.
Therefore the above relation does not constrain $\widehat{E}_F^a(z,z)$ at present.}
In order to see the 
qualitative behavior of the SIDIS cross sections, we fix those nonperturbative functions
so that the similar twist-3 mechanism associated with those functions
can reproduce the observed $A_N$ 
for $p^\uparrow p\to \pi X$\,\cite{E704,STAR}; the twist-3 cross section for
$p^\uparrow p\to \pi X$ also 
receive the SGP contributions analogous to the (A) and (B) terms of (\ref{twist3}),
although the additional parton distribution associated with the unpolarized nucleon participates.
For simplicity, we require that each of (A)- and (B)-type contributions can 
reproduce $A_N$ for $p^\uparrow p\to \pi X$: 
The authors of ref. \cite{QS99} used the ansatz $G_F^a(x,x)=K_a q_a(x)$ with a 
flavor-dependent constant $K_a$ to model the valence component of 
$G_F^a(x,x)$ using the unpolarized quark distribution $q_a(x)$, and found that 
the SGP contributions associated with the derivative 
${d\over dx}G_F^a(x,x)$ approximately 
reproduce $A_N$ observed in \cite{E704,STAR} with the value $K_u=-K_d=0.07$.

As was stated below (\ref{twist3frag}),
$\widehat{E}_F^a(z_1,z_2)$ does not have definite
symmetry property.  Here
we assume $2[(\partial/\partial z_1)\widehat{E}_F^a(z_1,z)]_{z_1=z}
=(d/dz)\widehat{E}_F^a(z,z)$ for simplicity.  With this assumption, 
the SGP contributions associated with the derivative of $\widehat{E}_F^a$
also reproduce
$A_N$ in \cite{E704,STAR}, using
the ansatz $\widehat{E}_F^a(z,z)=\widehat{K}_a D_a(z)$ with
$\widehat{K}_{u}^{\pi^+}=\widehat{K}_{d}^{\pi^-}=
-\widehat{K}_{\bar{d}}^{\pi^+}=-\widehat{K}_{\bar{u}}^{\pi^-}=-0.19$
and combining with the assumption for the transversity,  
$\delta q_d (x)=\Delta q_d (x)$ and $\delta q_u (x)=0.58 
\Delta q_u (x)$\,\cite{Koike03}.~\footnote{In \cite{Koike03}, 
we employed the same choice for the $d$-quark as above, 
but used $\widehat{K}_u^{\pi^+}=-0.11$ and
$\delta q_u (x)=\Delta q_u (x)$
for the $u$-quark, instead of the above values.  From the 
isospin and charge conjugation symmetry, however, one should have 
$\widehat{K}_{u}^{\pi^+}=\widehat{K}_{d}^{\pi^-}=
-\widehat{K}_{\bar{d}}^{\pi^+}=-\widehat{K}_{\bar{u}}^{\pi^-}$.
Combined with the new choice for the $u$-quark transversity, 
$\delta q_u (x)=0.58 \Delta q_u (x)$, 
the value $\widehat{K}_{u}^{\pi^+}=-0.19$ gives the same prediction
for $A_N$ as in \cite{Koike03}.}
In our estimate, 
we use GRV unpolarized parton distribution $q_a(x)$\,\cite{GRV98},
GRSV polarized parton distribution $\Delta q_a (x)$\,\cite{GRSV00}, 
and KKP pion fragmentation function $D_{a,g}(z)$\,\cite{KKP00}. 

To see the magnitude of each component in (\ref{azimuthA}) and (\ref{azimuthB}),
we calculate the azimuthal asymmetries normalized by the unpolarized cross section.
We define $\phi$-integrated azimuthal asymmetries as, setting $\Phi_S=\pi/2$,
\beq
\la 1 \ra_N \equiv {\sigma_0^{\rm A} \over \sigma_0^O},\qquad
\la\cos\phi\ra_N\equiv {\sigma_1^{\rm A}\over 2\sigma_0^O},\qquad 
\la\cos 2\phi\ra_N\equiv {\sigma^{\rm A}_2\over 2\sigma_0^O},
\label{ssaA}
\eeq
for the (A) contribution, and
\beq
\la 1\ra_N\equiv {\sigma_0^{\rm B}\over \sigma_0^O},\qquad
\la\sin(\Phi_S-\phi)\ra_N\equiv {\sigma_1^{\rm B}\over 2\sigma_0^O},\qquad 
\la\sin(\Phi_S-2\phi)\ra_N\equiv {\sigma^{\rm B}_2\over 2\sigma_0^O},
\label{ssaB}
\eeq
for the (B) contribution.
For the purpose of illustration, we choose two sets of 
kinematic variables in the region where our formula is valid.  
The first one is $S_{ep}=300$ GeV$^2$, $Q^2=100$ GeV$^2$, $x_{bj}=0.4$, 
which is close to the COMPASS kinematics.  Another one is
$S_{ep}=1000$ GeV$^2$, $Q^2=100$ GeV$^2$ and $x_{bj}=0.12$, which is in the region of
planned eRHIC experiment\,\cite{erhic}.  
Both sets give the same $\cosh\psi$ in (\ref{eq2.cosh}).

Fig.~3 shows the SSAs (\ref{ssaA}) for $ep^\uparrow\to e \pi^0 X$ from the (A) contribution.
One sees that 
$\la 1\ra_N$ can be as 
large as 5\% at $x_{bj}=0.4$, while $\la\cos\phi\ra_N$ and $\la\cos 2\phi\ra_N$
are much less than 1\%.  
For comparison we also showed the azimuthal asymmetries for the unpolarized
cross section (\ref{azimuthunpol}), $\la \cos n\phi\ra_O=\sigma^O_n/(2\sigma^O_0)$ ($n=1,\ 2$),
in the same figure.  $\la 1\ra_N$ is comparable to $\la \cos\phi\ra_O$.  
Here the derivative $(d/dx)G_F^a(x,x)$ together 
with $G_F(x,x)=K_a q_a(x)$, which explains rising $A_N$ for
$p^\uparrow p\to\pi X$ at
large $x_F$, causes similar rising $\la 1\ra_N$ toward $z_f\to 1$.  
We also note the SSA has smooth $q_T$-dependence. 
(Recall that the transverse momentum of the pion, $\ell_T$, is given by
$\ell_T=z_f q_T$ as (\ref{eq2.p_B}).)
 
Fig.~4 shows the azimuthal asymmetries (\ref{ssaB}) from the (B) contribution.
At both energies $S_{ep}= 300$ and 1000 GeV$^2$, these asymmetries
are much smaller than the (A) contribution of Fig.~3, although
the derivative $(d/dz)\widehat{E}_F^a(z,z)$ with
$\widehat{E}_F^a(z,z)=\widehat{K}_a D_a(z)$
causes rising behavior of SSA at large $z_f$.
From these calculations, 
the effect of the (B) contribution is negligible in $ep^\uparrow\to e\pi X$,
even if 
one fixes the strength of the nonperturbative functions so that
the corresponding (B)-type contribution 
solely reproduces $A_N$ for $p^\uparrow p\to\pi X$.

\section{Summary}

In this paper, we have studied the SSA for the large-$p_T$ pion production
in SIDIS $ep^\uparrow\to e\pi X$.
We derived the twist-3 cross section formula, 
focussing on the the SGP contributions associated with
the derivative of 
the twist-3 distribution $G_F$ for the nucleon, and with the derivative of the twist-3 
fragmentation function $\widehat{E}_F$ for the pion.
This approximation should be valid in the region where $x_{bj}$ or $z_f$ is large. 
In estimating the impact of each contribution,
we fixed the overall 
strengths of $G_F$ and $\widehat{E}_F$
so that each of them independently
reproduces $A_N$ in $p^\uparrow p\to\pi X$, and found that the 
$\widehat{E}_F$ contribution is negligible compared to the $G_F$ contribution. 
A full calculation including the nonderivative terms of the SGP contributions
as well as the SFP and HP contributions will be reported elsewhere.  

\vspace{0.5cm}

\noindent
{\it Note added:}

After the submission of this work, the preprint \cite{JQVY06SIDIS} appeared.  
The authors of this reference have performed the twist-3 calculation for
the $qq$-contribution (left figure of
Fig. 1(a)) to $\sigma_0^{\rm A}$ term in (\ref{azimuthA}), including all the SGP (both derivative and nonderivative terms) 
and HP contributions, and demonstrated 
its equivalence with the Sivers effect in the small $p_T$ region.
Their result for the derivative term of the SGP contribution agrees with ours.

\begin{figure}[t!]
\begin{center}
\vspace*{0.8cm}
\hspace*{-5mm}
\epsfig{figure=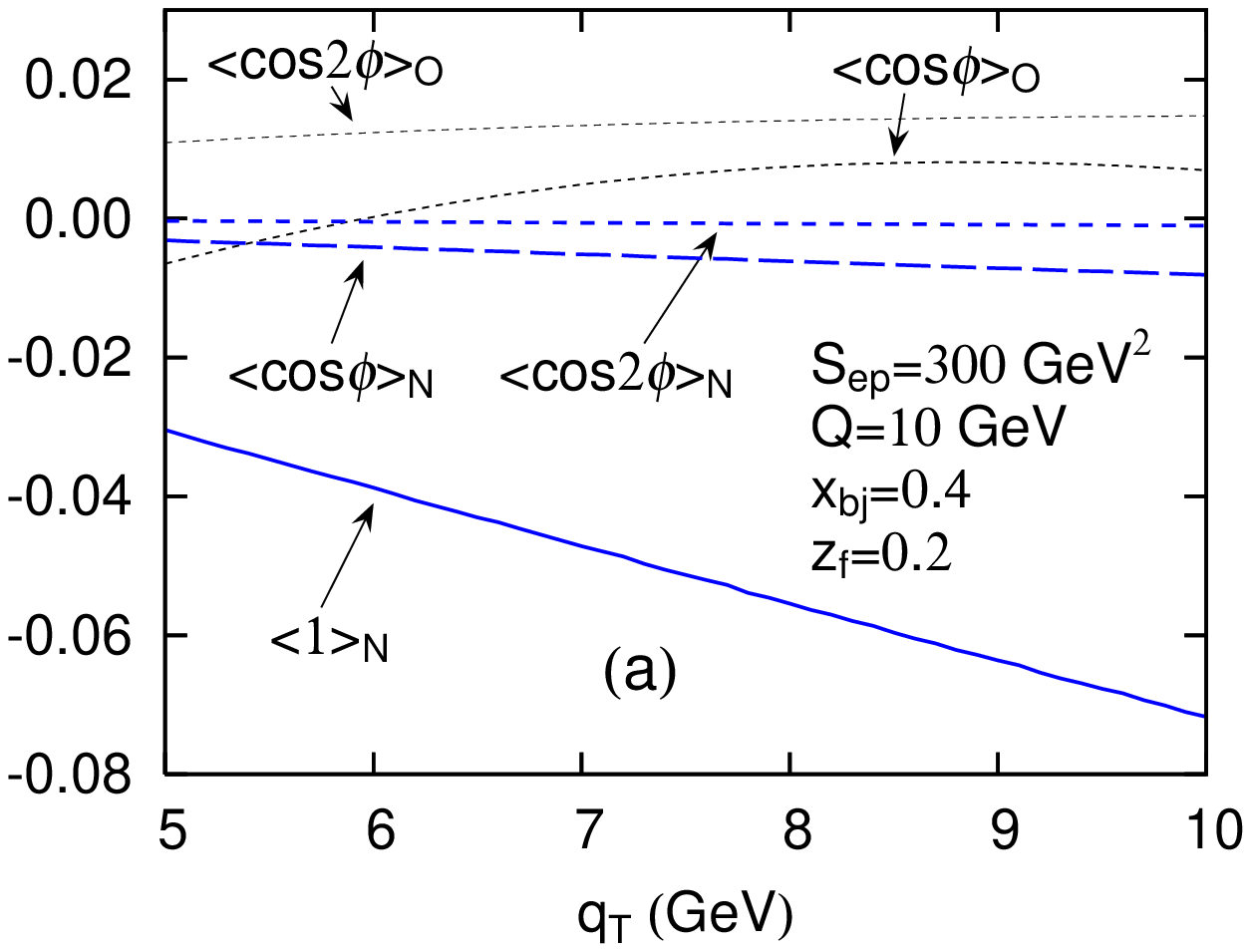,width=0.45\textwidth}
\hspace*{10mm}
\epsfig{figure=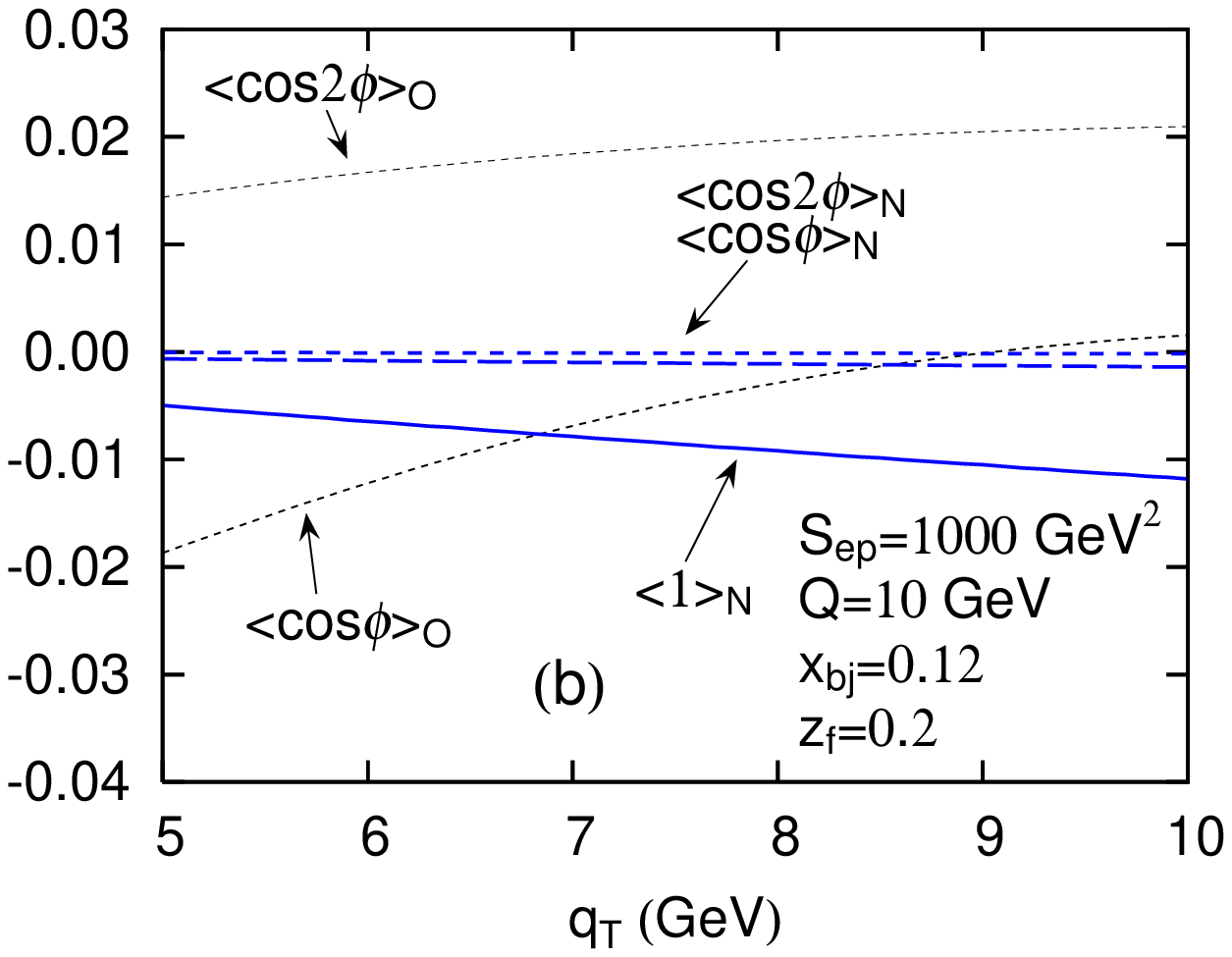,width=0.45\textwidth}

\vspace*{0.5cm}
\hspace*{-5mm}
\epsfig{figure=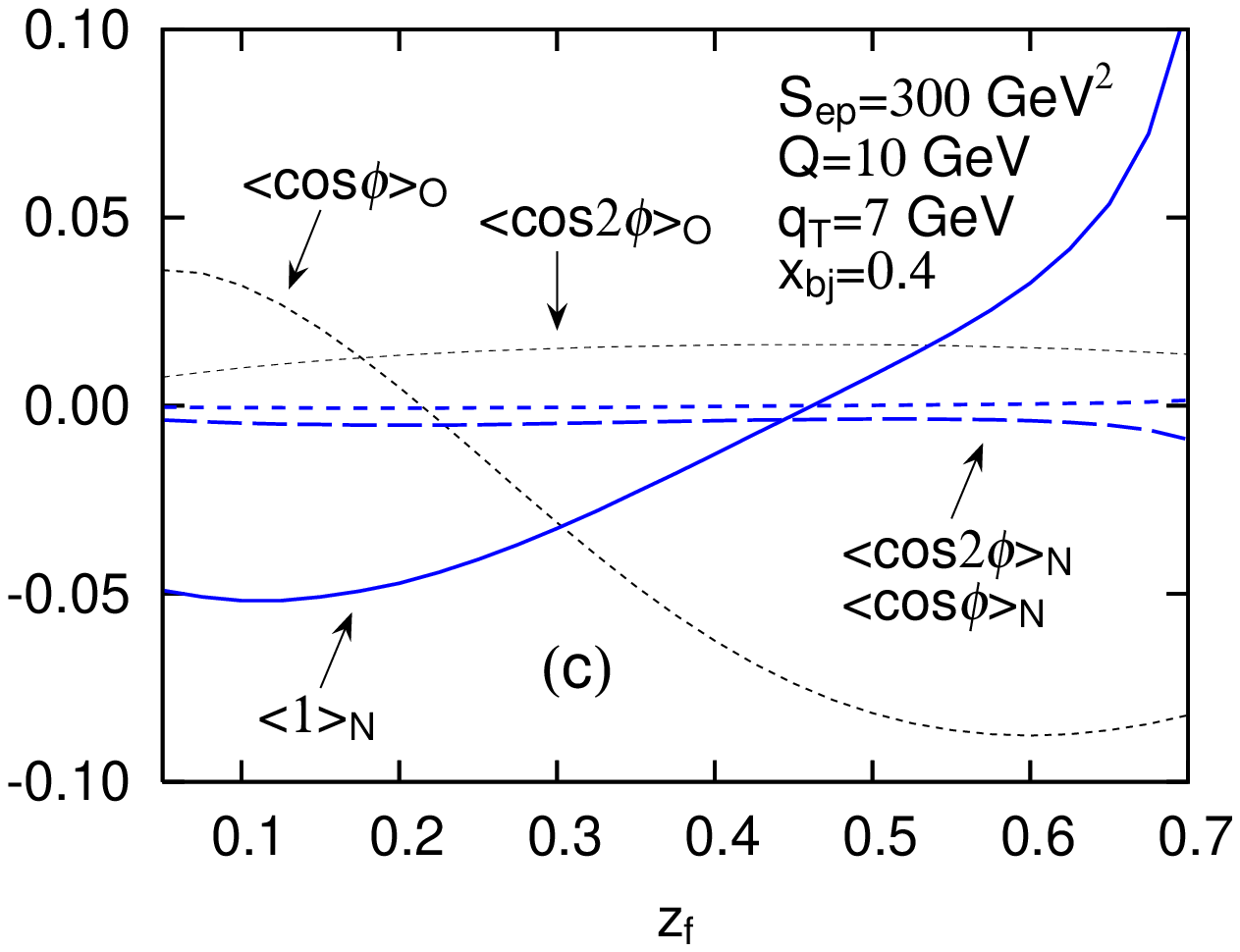,width=0.45\textwidth}
\hspace*{7.5mm}
\vspace{3mm}
\epsfig{figure=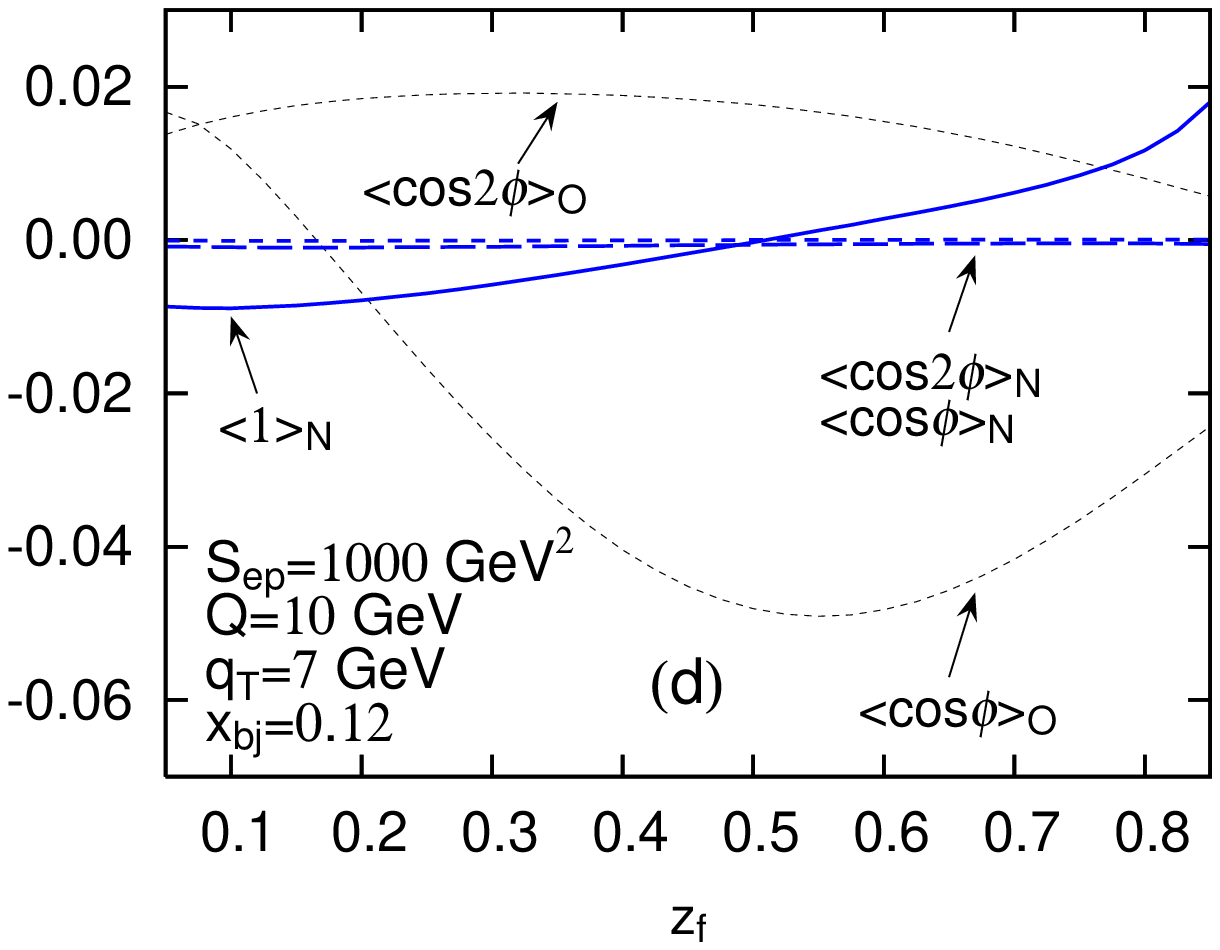,width=0.44\textwidth}
\end{center}
\vspace*{-.5cm}
\caption{
The azimuthal asymmetries for the $G_F$ contribution to
 $ep^\uparrow\to e\pi^0 X$
at two energies.  (a) and (b) show the $q_T$-dependence, while
(c) and (d) show the $z_f$ dependence.
\label{fig3}}
\vspace*{0.cm}
\end{figure}

\begin{figure}[t!]
\begin{center}
\vspace*{0.8cm}
\hspace*{-5mm}
\epsfig{figure=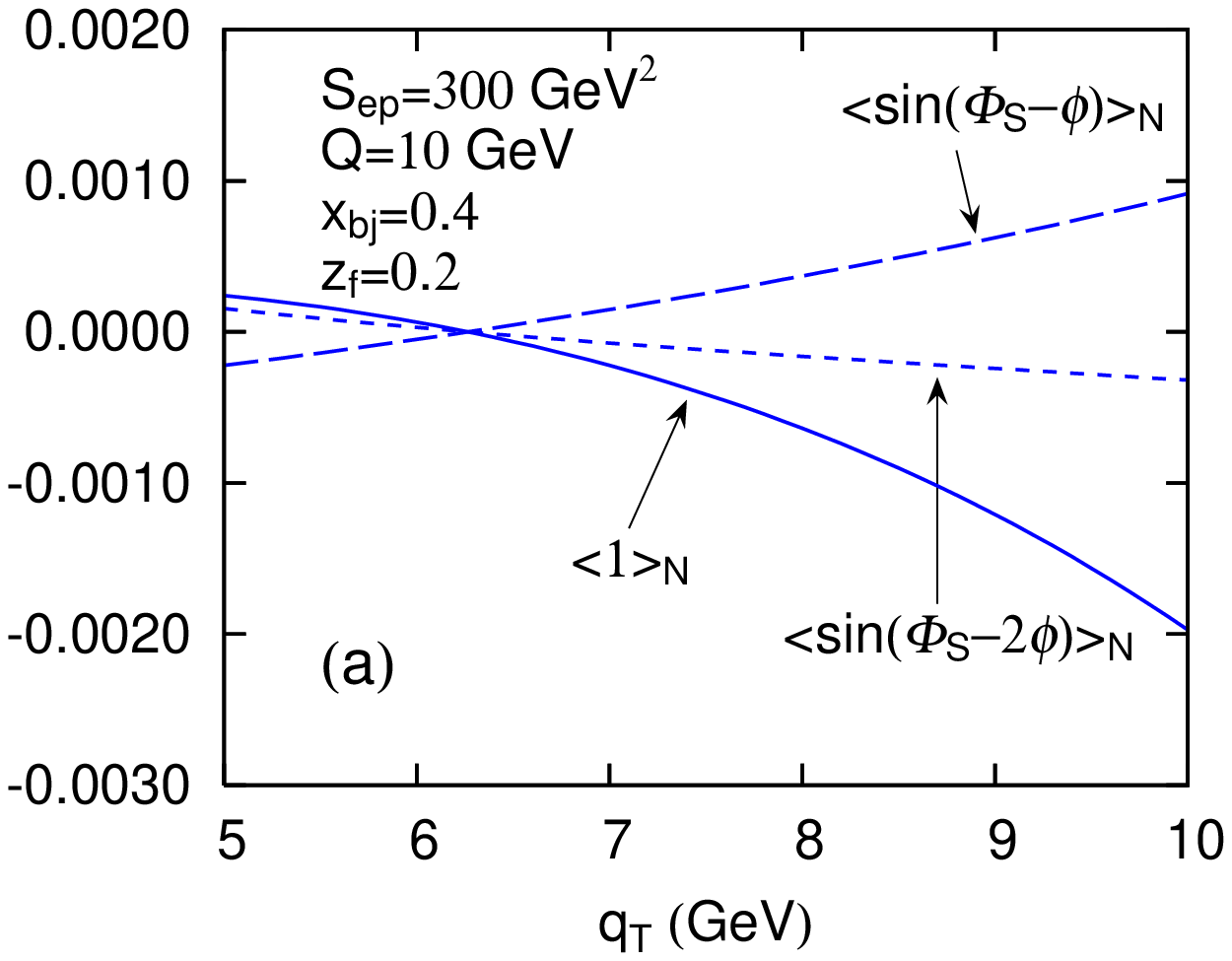,width=0.45\textwidth}
\hspace*{10mm}
\epsfig{figure=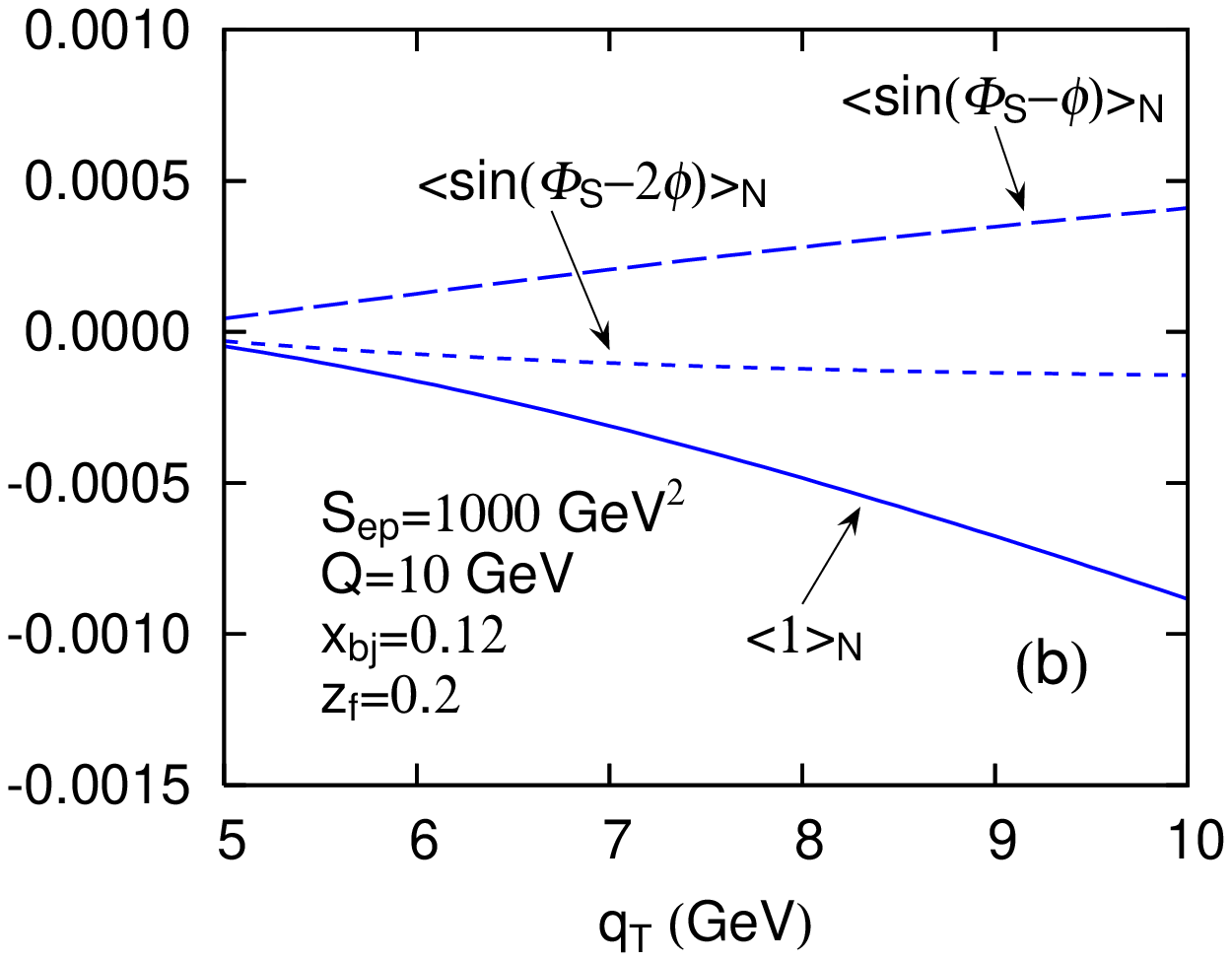,width=0.45\textwidth}

\vspace*{0.5cm}
\hspace*{-5mm}
\epsfig{figure=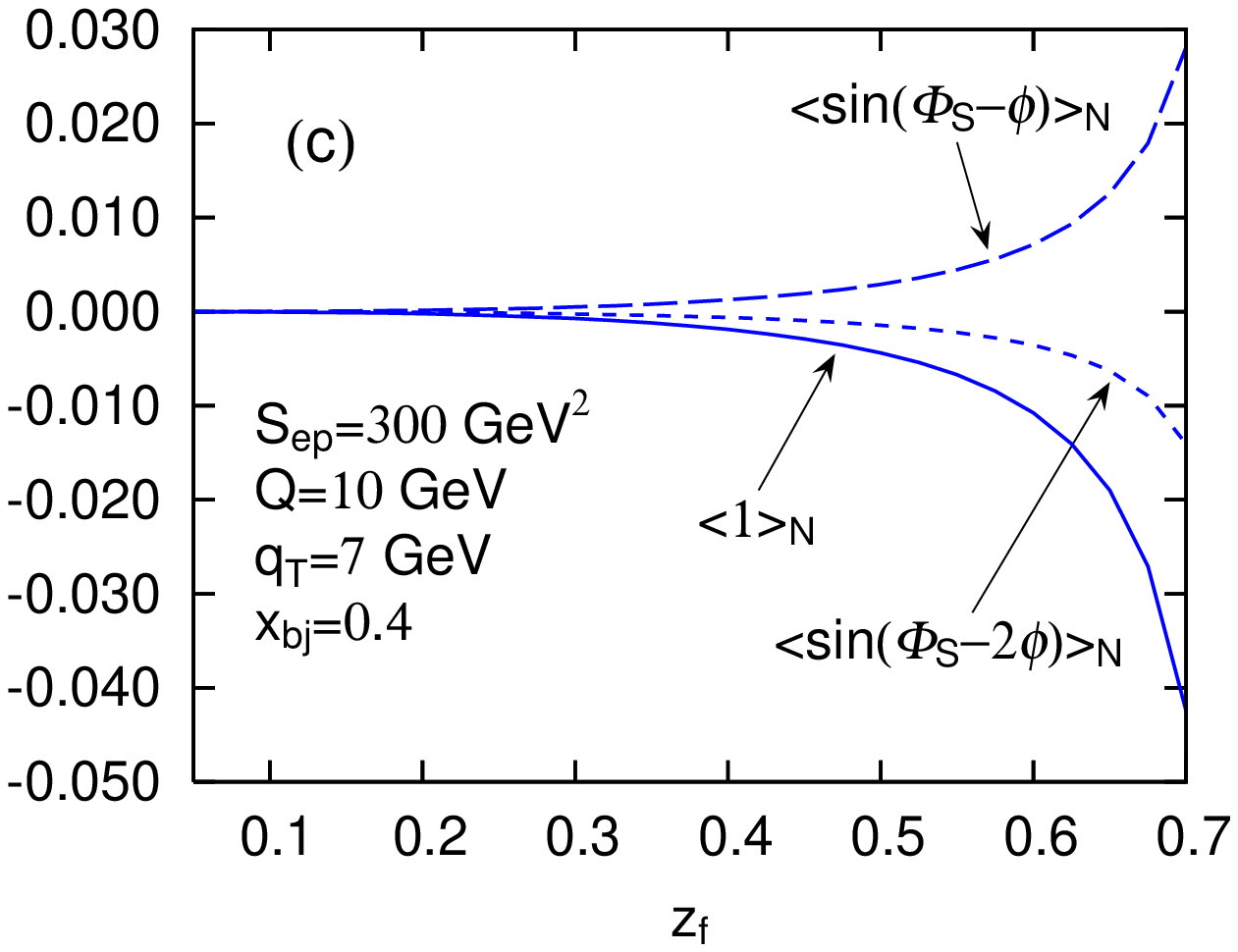,width=0.45\textwidth}
\hspace*{7.5mm}
\vspace{3mm}
\epsfig{figure=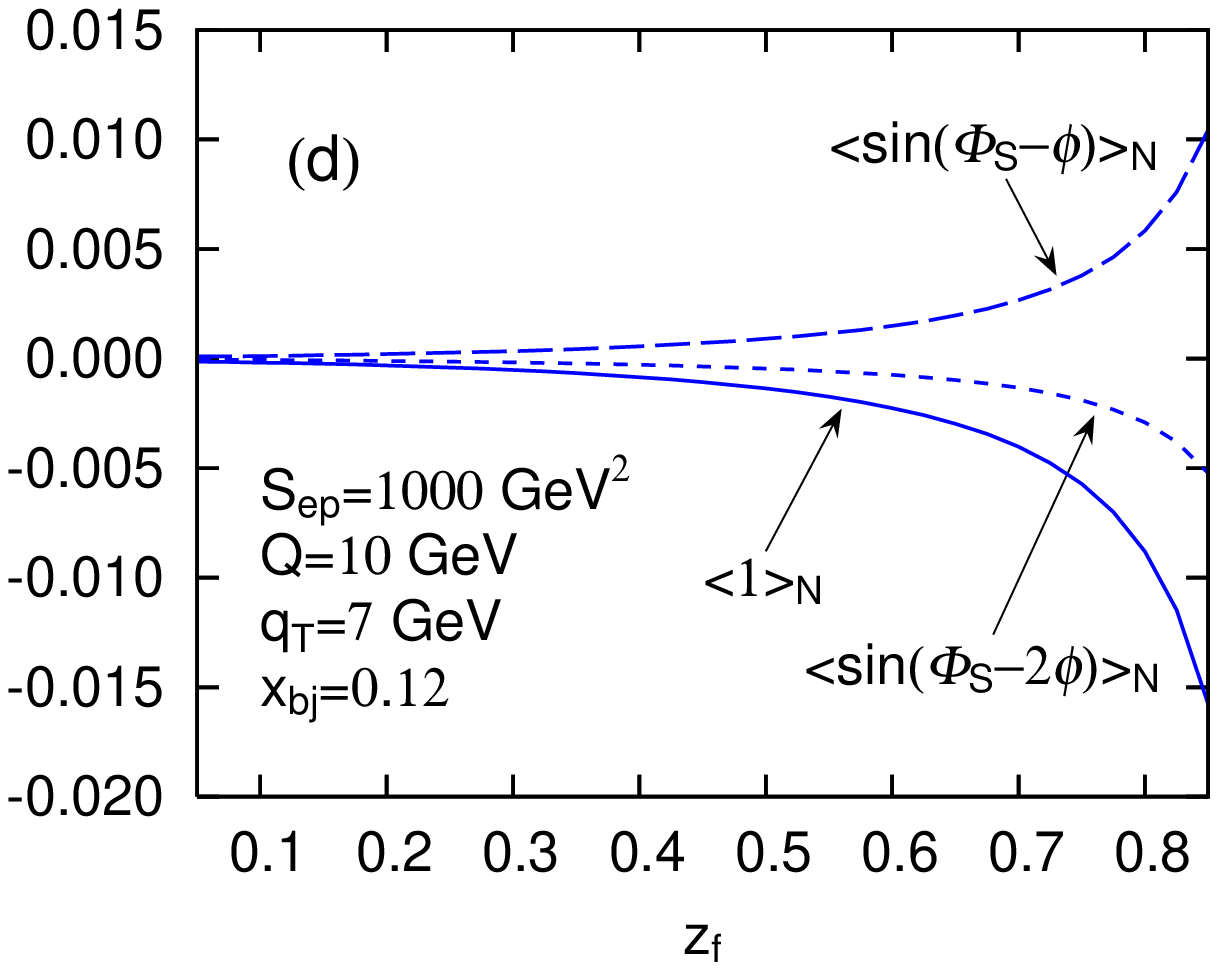,width=0.44\textwidth}
\end{center}
\vspace*{-.5cm}
\caption{
The azimuthal asymmetries for the $\widehat{E}_F$ contribution to
 $ep^\uparrow\to e\pi^0 X$
at two energies.  (a) and (b) show the $q_T$-dependence, while
(c) and (d) show the $z_f$ dependence.
\label{fig4}}
\vspace*{0.cm}
\end{figure}

\section*{Acknowledgements}
The work of K.T. was 
supported by the Grant-in-Aid for Scientific Research No. C-16540266.

\appendix

\section*{Appendix: Relation between twist-3 distributions}

In the main text, we have discussed the 
relations (\ref{FDvector})-(\ref{gt}) between the $D$-type and $F$-type 
twist-3 correlation functions for 
the transversely polarized nucleon. This appendix is devoted to demonstrate
that 
$\gt(x)$ of (\ref{gt})  
can be expressed in terms of the $F$-type 
twist-3 correlation functions 
and the twist-2 quark helicity distibution $\Delta q(x)$ for the nucleon.
Such relation involves the components with different number of partons, and can be
obtained as a consequense of constraints from the equations of motion and Lorentz invariance.

We start with the well-known definition of the spin-dependent quark distributions
of the nucleon\,\cite{Jaffe:1991ra,KT99},
\begin{equation}
\int { d\lambda  \over 2 \pi } e^{i\lambda x} \langle p S |
  \bar{\psi}(0) [0, \lambda n]\gamma^{\mu}\gamma_5 
  \psi(\lambda n) | p S \rangle
 = 2 M_N \left[ \Delta q(x) p^{\mu}(S\cdot n)
   + g_{T}(x) S_{\perp}^{\mu} 
\right]+\cdots, \label{eq:axialv}
\end{equation}
where $\Delta q(x)$ and $g_T (x)$ are the twist-2 and -3 quark distributions
for the longitudinally and transversely polarized nucleon, respectively,
and 
``$\cdots$'' denotes Lorentz structure of twist higher than three.
Using the QCD equations of motion, it can be shown\,\cite{Ratcliffe:1985mp} that
$g_T(x)$ can be expressed in terms of the $D$-type functions as
\begin{equation}
g_T(x)=-\frac{1}{4x}\int dx' \left[ \GDt(x , x' )
+\GDt(x' , x )+ G_D (x , x' )
- G_D (x' , x )\right].
\label{rat}
\end{equation}
Substituting the relations (\ref{FDvector}) and (\ref{FDrelation}) 
into (\ref{rat}) and using the symmetrty properties (\ref{sym}),
we get
\begin{equation}
g_T(x)= -\frac{1}{2x}\left( \gt(x) 
+ P\int_{-1}^{1} dx' \frac{G_F (x , x' ) + \GFt(x , x' )}{x -x'} \right).
\label{rat2}
\end{equation}
We note that this result has been obtained restricting the quark and gluon fields strictly on
the lightcone. As is well-known, however, 
similar result can be obtained using the operator product
expansion, where the fields are not 
restricted on the lightcone and thus the constraints from Lorentz 
invariance are fully taken into account.
Here it is convenient to empoy the nonlocal version of the operator product expansion using the 
operator dentity\,\cite{Balitsky:1987bk,KT99}
\begin{eqnarray}
  \lefteqn{ z_{\mu} \left(
     \frac{\partial}{\partial z_{\mu}}
        \bar{\psi}(0) \gamma^{\sigma} \gamma_5 [0, z] \psi( z )
  -  \frac{\partial}{\partial z_{\sigma}}
        \bar{\psi}(0) \gamma^{\mu} \gamma_5 
       [0, z] \psi( z ) \right)}\nonumber\\
  &=& \int^1_0 dt \bar{\psi}(0) [0, tz] \sslash{z} \left\{
          i \gamma_5 \left( t - \frac{1}{2} \right) g F^{\sigma\rho}(tz)
           z_{\rho} - \frac{1}{2} 
g \tilde{F}^{\sigma\rho}(tz) z_{\rho} \right\} 
     [tz , z] \psi ( z ) \nonumber\\
   & & + \, \left[ \bar{\psi}(0)
           \gamma_5 \sigma^{\sigma\rho} z_{\rho} [ 0 , z]
i \lslash{D} 
\psi (z)  - \bar{\psi}(0)
i \overleftarrow{\lslash{D}} 
          \gamma_5 \sigma^{\sigma\rho} z_{\rho} [ 0 ,z] \psi ( z )
                \right],
\label{eq:identity}
\end{eqnarray}
where $z_\mu$ is not restricted on the lightcone. 
This identity is exact up to the irrelevant terms, i.e.,
the terms proportional to quark mass,
the twist-4 ($O(z^2)$) contributions, 
and the total derivatives. In the lightcone limit $z^2 \rightarrow 0$,
the nucleon matrix element of operators on both sides can be expressed in terms of the 
distribution functions defined in (\ref{eq:axialv}):
\begin{eqnarray}
 \langle p\ S |
  \bar{\psi}(0) [0, z]\gamma^{\mu}\gamma_5 
  \psi(z) | p\ S \rangle
 &=& 2M_N \int_{-1}^{1} dx e^{-ixp\cdot z} 
\left( p^{\mu}\frac{S\cdot z}{p\cdot z}\left[ \Delta q(x) -g_T (x) + O(z^2) \right]\right.
 \nonumber\\
 & +& \left. S^{\mu}\left[ g_{T}(x) + O(z^2) 
\right]\right) +\cdots, \label{eq:axialv2}
\end{eqnarray}
and the $F$-type distributions of (\ref{twist3distr}); note that the matrix element 
of the last line of (\ref{eq:identity})
vanishes using the equations of motion.
Thus the matrix element of (\ref{eq:identity}) yields the differential equation
\begin{eqnarray}
\lefteqn{x\frac{dg_T(x)}{dx}+\Delta q(x) }\nonumber\\
&&=-\frac{1}{2} P\int_{-1}^{1} dx' \frac{1}{x-x'}
\left\{ \left(\frac{\partial}{\partial x}+\frac{\partial}{\partial x'}\right)
G_F(x, x' )+
\left(\frac{\partial}{\partial x}-\frac{\partial}{\partial x'}\right)
\GFt(x, x' )
\right\},
\label{diffeq}
\end{eqnarray}
where we have used the symmetry relation (\ref{sym}).
The solution of this equation with the boundary condition, $g_T (x)=0$ for $|x|>1$, reads
\begin{eqnarray}
g_T(x)&=&\int^{\epsilon(x)}_x \frac{dx_1}{x_1}\left[  \Delta q(x_1 )
+ \frac{1}{2} P\int_{-1}^{1} dx_2 \frac{1}{x_1 -x_2}
\left\{\left(\frac{\partial}{\partial x_1}+\frac{\partial}{\partial x_2}\right)
G_F(x_1 , x_2 ) 
\right. \right. \nonumber \\
&&\left. \left.
+\left(\frac{\partial}{\partial x_1}-\frac{\partial}{\partial x_2}\right)
\GFt(x_1 , x_2 )
\right\}\right],
\label{sol}
\end{eqnarray}
where $-1 \le x \le 1$, and the first term on the RHS 
gives the Wandzura-Wilczek part for $g_T(x)$\,\cite{KT99}.
It is straightforward to see that the moments $\int dx x^n g_T (x)$ using this solution 
reproduce
the results obtained by the standard local operator expansion.
Comparing (\ref{rat2}) with (\ref{sol}), we find that $\gt(x)$ is expressed  
in terms of $G_F (x_1 , x_2 )$, $\GFt (x_1 , x_2 )$, and $\Delta q(x)$,
and get
\begin{eqnarray}
\gt(x) =-  x  \int^{\epsilon(x)}_x dx_1 \left[  \frac{2\Delta q(x_1 )}{x_1}
+ \frac{1}{x_1^2}P\int_{-1}^{1} dx_2 \left\{
\frac{G_F (x_1 , x_2 )}{x_1 -x_2}
+ 
\left( 3x_1 -x_2 
\right)\frac{\GFt(x_1 , x_2 )}{(x_1 -x_2 )^2}
\right\} \right].
\label{gtsol}
\end{eqnarray}
Combined with (\ref{FDrelation}), 
this result shows that $\GDt(x_1 , x_2)$ contains the contribution
from the twist-2 operators corresponding 
to the Wandzura-Wilczek part, and the genuine twist-3 part 
of the $D$-type distributions $G_D (x_1 , x_2 )$, $\GDt (x_1 , x_2 )$ can be completely expressed  
in terms of the $F$-type distributions $G_F (x_1 , x_2 )$ and $\GFt (x_1 , x_2 )$.


\end{document}